\documentclass[a4paper,11pt]{article}
\pdfoutput=1 
\usepackage{jcappub} 
\usepackage[normalem]{ulem}

\graphicspath{{.}{./figures/}}
\usepackage[T1]{fontenc} 
\let\vec\mathbf

\usepackage[mathletters]{ucs}
\usepackage[utf8x]{inputenc}

\newcommand{\Fermi}{{\it Fermi}}

\begin{document}\

\begin{flushright}
LAPTH-012/18
\end{flushright}

\title{ Gamma-ray spectral modulations of Galactic pulsars caused by photon-ALPs mixing}

\author[a]{Jhilik Majumdar,}
\author[b]{Francesca Calore,}
\author[a]{Dieter Horns}
\affiliation[a]{Institut f\"ur Experimentalphysik, University of Hamburg, \\Hamburg,Germany}
\affiliation[b]{Univ. Grenoble Alpes, USMB, CNRS, LAPTh, F-74940 Annecy, France}
\emailAdd{Jhilik.majumdar@desy.de}

\abstract{
%
	  Well-motivated extensions of the standard model predict ultra-light and fundamental pseudo-scalar particles (e.g., axions or axion-like particles: ALPs).  
%
%
Similarly to the Primakoff-effect for axions, ALPs can mix with photons and 
consequently be searched for in laboratory experiments and with astrophysical observations.
Here, we search for energy-dependent modulations of high-energy gamma-ray spectra that are tell-tale signatures of photon-ALPs mixing. 
To this end, we analyze the data recorded with the \Fermi-LAT from
Galactic pulsars selected to have  a line of sight crossing  spiral arms at a large pitch angle. 
The large-scale Galactic magnetic field traces the shape of 
spiral arms, such that a sizable  photon-ALP conversion probability is expected for the sources considered.
In nine years of \Fermi-LAT data, we  detect significant spectral features in the selected source-sample consistent with photon-ALPs oscillation with a combined statistical significance of 5.52~$\sigma$. 
Notably, sources with neighboring 
lines of sight share similar spectral features. 
From a common fit to all sources, we determine the most-likely parameters for mass $m_a$ and coupling
$g_{a\gamma\gamma}$ to be $m_a=(3.6 \substack{+0.5_ \mathrm{stat.}\\-0.2_ \mathrm{stat.}}\pm 0.2_\mathrm{syst.} )$~neV and $g_{a\gamma\gamma}=(2.3\substack{+0.3_ \mathrm{stat.}\\-0.4_ \mathrm{stat.}}\pm 0.4_\mathrm{syst.})\times 10^{-10}$~GeV$^{-1}$.
In the error budget, we consider  instrumental effects, scaling of the adopted Galactic magnetic field model ($\pm~20~\%$), and uncertainties on the  distance of individual sources.
 We note that an astrophysical interpretation of the detected modulation is
not obvious.}

\maketitle

\section{Introduction}
\label{sec:intro}
Among the possible extensions of the standard model of particle physics (SM), 
the \textit{axion} is an elegant and necessary addition which was initially suggested to cure the strong CP problem \cite{peccei_cp_1977,wilczek_problem_1978}. The axion has the interesting property to mix 
with photons via the Primakoff process, leading to an effective Lagrangian term of the form:
\begin{equation}
\label{eq:axionLagr}
\mathcal{L} \supset - \frac{1}{4}  g_{a\gamma\gamma} F_{\mu\nu} \tilde{F}^{\mu\nu} a = g_{a\gamma\gamma} \, \vec{E}\cdot\vec{B} \, a \, ,
\end{equation}
where, $a$ is the axion field with mass $m_{a}$, $F_{\mu\nu}$ is the
electromagnetic field-strength tensor, $\tilde{F}^{\mu\nu} =
\frac{1}{2}\varepsilon_{\mu\nu\rho\sigma}F^{\rho\sigma} $ its dual,
$g_{a\gamma\gamma}$ is the photon-axion coupling constant which leads to
 oscillations between photon and axion states, e.g.,  in the
presence of external transversal magnetic fields \cite{raffelt_mixing_1988,anselm_experimental_1988}. 
For axions, $g_{a\gamma\gamma}$ is related to the axion mass and 
to the energy scale  at which the Peccei-Quinn symmetry is
broken. More generally, axion-like particles (ALPs) are predicted in several
string-theory-motivated extensions of the SM~\cite{anselm_second_1982,dias_quest_2014,cicoli_type_2012}.
In these extensions, the ALPs are not necessarily related to the strong CP
problem. Nevertheless, ALPs  could couple to  photons as expressed in eq.~\ref{eq:axionLagr}. Differently from axions, mass and coupling constant of ALPs are  not
necessarily related to each other. 

 Various astrophysical observations have been suggested and used to test the
ALPs hypothesis. Common to many approaches is photon-ALPs mixing
in external magnetic fields (see e.g.~\cite{horns_gamma_2016} for a review on 
relevant gamma-ray observations) or plasma-related contributions 
to the $\vec{E}\cdot\vec{B}$ term in eq.~\ref{eq:axionLagr} (e.g., helioscopes \cite{sikivie_experimental_1984} and
prompt gamma-rays from supernova explosions \cite{brockway_sn_1996}).\\
   The photon-ALPs oscillation (see eq.~\ref{eq:axionLagr})    
   is efficient at energies larger than a critical
photon energy $E_c$ as, e.g., given by ($\hbar=c=1$) \cite{de_angelis_evidence_2007}: 
\begin{equation}
\label{eqn:ecrit}
 E_{c}\simeq 2.5~\mathrm{GeV}  \frac{|m_{a}^2-\omega_{Pl}^2|}{1\,\mathrm{neV}} \left(\frac{B_{\perp}}{\mu\rm G}\right)^{-1} \left(\frac{g_{a\gamma\gamma}}{10^{-11}~\mathrm{GeV}^{-1}}\right)^{-1} \, , 
 \end{equation}
 where $\omega_{pl}=0.03~\mathrm{neV}~n_e^{1/2}$  is the plasma frequency
 in a medium with electron density $n_e$ in electrons per $\mathrm{cm}^{-3}$, $B_{\perp}$ is the transversal magnetic field and $g_{a\gamma\gamma}$ is the photon-ALPs coupling constant. While for $E_\gamma\gg E_c$ the conversion probability is independent of the
 photon-ALPs mixing angle and photon energy, it gets inefficient for energies $E_\gamma \ll E_c$.
 At energies close to the critical energy, the mixing depends on the energy and leads to
 spectral features in observed spectra. For large scale magnetic
 fields, the oscillation length is given as
\cite{mirizzi_stochastic_2009}
\begin{equation}
\label{eqn:losc}
l_\mathrm{osc}= 32~\mathrm{kpc}\sqrt{1+(E_c/E_\gamma)^2} \left(\frac{B_\perp}{\mu\mathrm{G}}\right)^{-1}
                                \left(\frac{g_{a\gamma\gamma}}{10^{-11}~\mathrm{GeV}^{-1}}\right)^{-1}, 
\end{equation} 
 which implies that, for typical parameters in the Galaxy, 
 oscillations can be relevant on Galactic scales at GeV energies.
 
 In previous works related to ALPs signatures in gamma-ray spectra, there have been various approaches and sources considered, e.g., anomalous transparency of extra-galactic sources 
 \cite{simet_milky_2008,de_angelis_photon_2009,horns_indications_2012,kohri_axion-like_2017},
 as well as searches for disappearance effects from extra-galactic objects in 
 the magnetic field of a galaxy cluster  \cite{ajello_search_2016},
 and in intergalactic space \cite{abramowski_constraints_2013}. For further discussion see section~\ref{sec:summary}.

\begin{figure}
\setlength{\unitlength}{.9cm}
\begin{center}
\begin{picture}(10,11)
 \put(-3.5,0){\includegraphics[width=15cm]{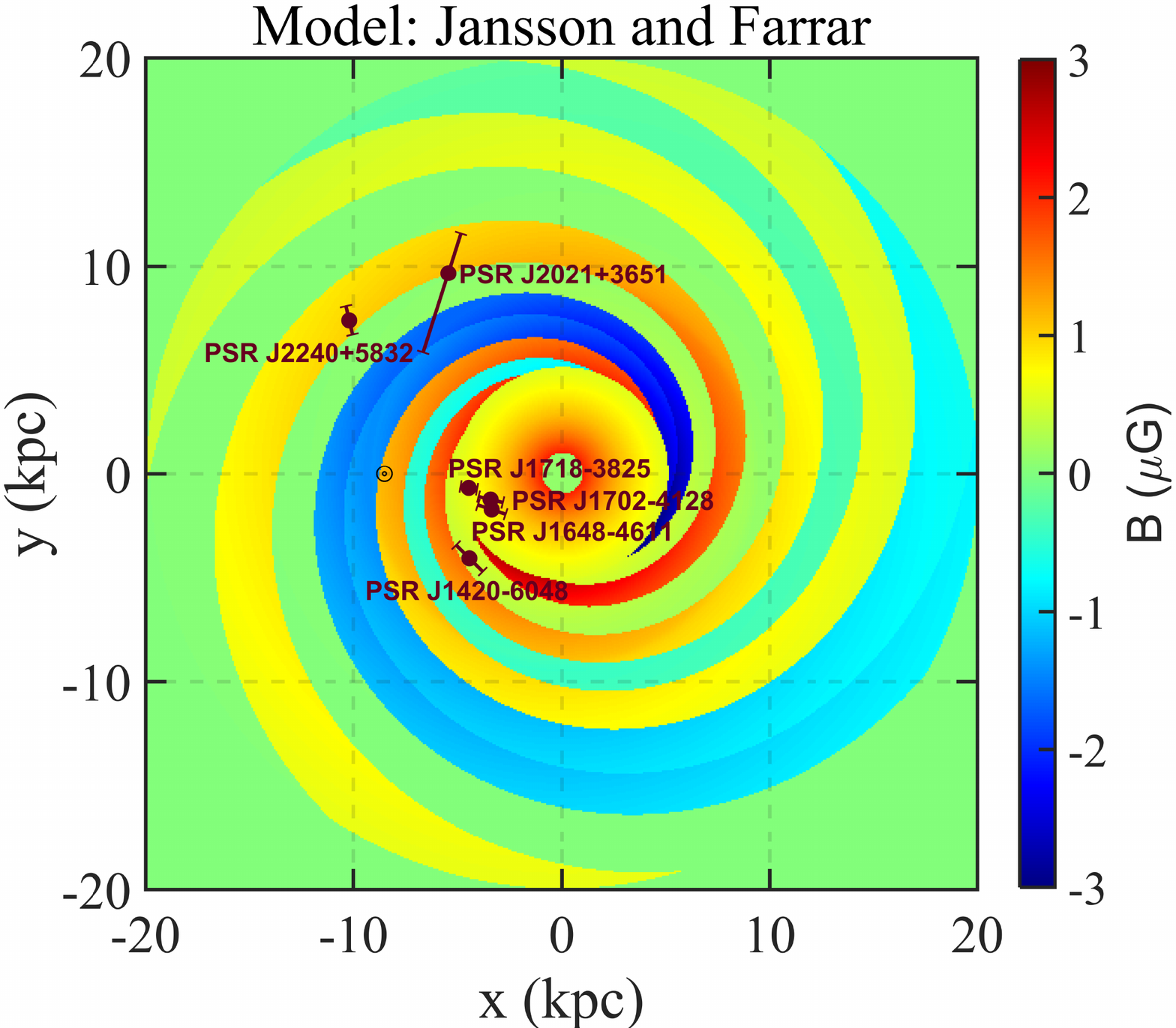}}
 \end{picture}
\end{center}

\caption{Source positions in the Galactic plane with the Jansson and Farrar model~\cite{jansson_new_2012} magnetic field strength indicated by the color scale. Both,
	PSR~J2021+3651 and PSR~J2240+5832 are located close to the fifth spiral arm, while 
	PSR~J1420-6048, PSR~J1648-4611, PSR~J1702-4128, and PSR~J1717-3825 are in the direction towards
	the Galactic center. Error bars at the source positions mark the 
	uncertainties on heliocentric distances. The position of the sun (at $x=-8.5$~kpc) is marked as well. 
	\label{fig:source_location}}

\end{figure}

In this article,  we investigate for the first time photon-ALPs oscillation features in the disappearance channel from Galactic gamma-ray pulsars. We use publicly available \Fermi-LAT data for six bright gamma-ray pulsars and search for  spectral irregularities that might be induced by photon-ALPs oscillations in the regular Galactic magnetic field. The nearby Vela pulsar is used as a reference source: since Vela's distance is much smaller than the oscillation length, no spectral modulations are expected.

The manuscript is organised as follows: in section~\ref{section:source_selection}, we discuss the selection of  six bright 
gamma-ray pulsars used for the present analysis. In sections~\ref{section:data_analysis} and~\ref{section:systematics},  we present the analysis of \Fermi-LAT data  and the assessment of systematic errors related to instrumental effects. In section \ref{section:results}, we present the results of the spectral fits performed with and without photon-ALPs mixing. In section~\ref{sec:summary}, we summarize and discuss our results.

 \begin{figure}
\setlength{\unitlength}{.9cm}
\begin{center}
\begin{picture}(10,12)
 \put(-3.5,0){\includegraphics[width=15cm]{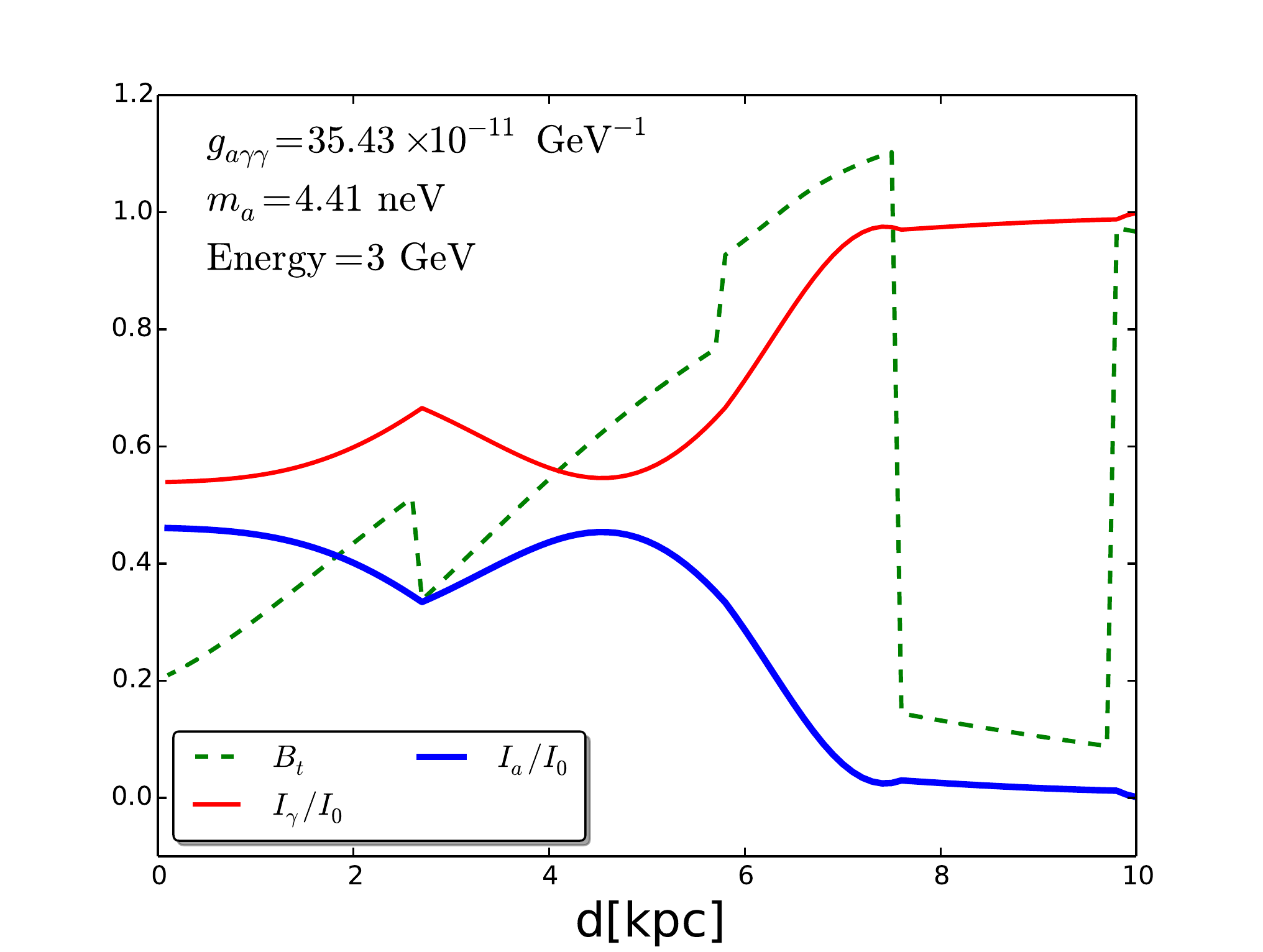}}
 \end{picture}
\end{center}
\caption{Photon (red thin line) and ALPs (blue thick line) intensity 
	along the line of sight towards PSR~J2021+3651.  
	 The green dashed line marks the transversal magnetic field. \label{fig:los} 
	 }

\end{figure} 

\section{Data analysis}
\subsection{Source selection}
\label{section:source_selection}
The shortest oscillation length (see eq.~\ref{eqn:losc}) and therefore the strongest effect is
expected for sources located at a large distance and along a line of sight with large $B_\perp$.
The source population with the best
determined distances are pulsars. Pulsars  have been observed in the entire Galaxy. We have
selected from the \Fermi~pulsar catalog the brightest pulsars with known distances
and lines of sight that traverse spiral arms at large pitch angles \cite{majumdar_modulations_2017,majumdar_spectral_2017}. 
 To date, about 160 gamma-ray pulsars have been observed with \Fermi-LAT 
 \cite{abdo_second_2013}. After applying the selection criteria (known distance and located at a favorable lines of sight), we have chosen  the resulting six brightest gamma-ray  pulsars from 
 the second \Fermi-LAT pulsar catalog (see table~\ref{tab:selection}). 
 The positions of the pulsars including uncertainties on their heliocentric distance are marked in figure~\ref{fig:source_location}.
 \\
    All six pulsars are rotation powered and fairly young. PSR J1420-6048 at a distance of ($5.7\pm0.9$) kpc (the distance is estimated from dispersion measure of radio-timing data \cite{weltevrede_gamma-ray_2010-1}) is a 68 ms pulsar  in the Kookaburra nebula which has been 
     extensively studied in X-ray, radio, and infrared \cite{roberts_multiwavelength_2001}.  PSR J1648-4611 at a distance of ($4.9\pm0.7$)~kpc (the distance is estimated from the dispersion measure given in \cite{kramer_parkes_2003} and using the electron distribution model \cite{cordes_ne2001.i._2002}) is tentatively associated with a very-high-energy (VHE) gamma-ray source observed with HESS \cite{abramowski_discovery_2012} in the vicinity of the massive stellar cluster Westerlund 1. 
     The pulsars PSR  J1702-4128 at  a distance of ($4.7\pm0.6$)~kpc and  PSR J1718-3825   at a distance of ($3.6\pm0.4$)~kpc
      have been associated with   pulsar wind nebulae  at VHE energies \cite{aharonian_discovery_2007,aharonian_hess_2008}.
     PSR J2021+3651 (distance of $10\substack{+2 \\ -4}$~kpc from dispersion measure and
     X-ray absorption \cite{hessels_observations_2004}) is a 17 kyr pulsar detected in radio, X-rays, and gamma rays (possibly associated with
     VER~J2019+368 \cite{aliu_spatially_2014}). This object resembles the Vela pulsar.
      We note that a recent X-ray absorption study \cite{kirichenko_optical_2015} favors a smaller distance of
     $1.8\substack{+1.7 \\  -1.4}$~kpc.
     The recently discovered northern-hemisphere pulsar PSR~J2240+5832 is located in an outer spiral arm similar to PSR~J2021+3651 at a distance
     of $(7.7\pm0.7)$~kpc \cite{theureau_psrs_2011}.  
    \\
    These objects are located at low Galactic latitude so that the emitted  photons traverse Galactic spiral arms (see figure~\ref{fig:source_location}).
       In order to estimate systematic uncertainties on the observed spectrum we use the Vela pulsar as a reference source. This pulsar is sufficiently close
       to determine a geometrical parallax distance of $294\substack{+76\\-50}$~pc \cite{caraveo_distance_2001}.
       Given Vela's apparent brightness, the gamma-ray spectrum is very well measured and does not show any spectral distortion. 
       To derive the systematic
       uncertainties, we use a similar technique to the \Fermi-LAT Pass 7 data analysis in~\cite{ackermann_fermi_2012}, see also section~\ref{section:systematics}. 
\begin{figure}
\setlength{\unitlength}{.9cm}
\begin{center}
\begin{picture}(10,8)
 \put(-0.5,0){\includegraphics[width=10cm]{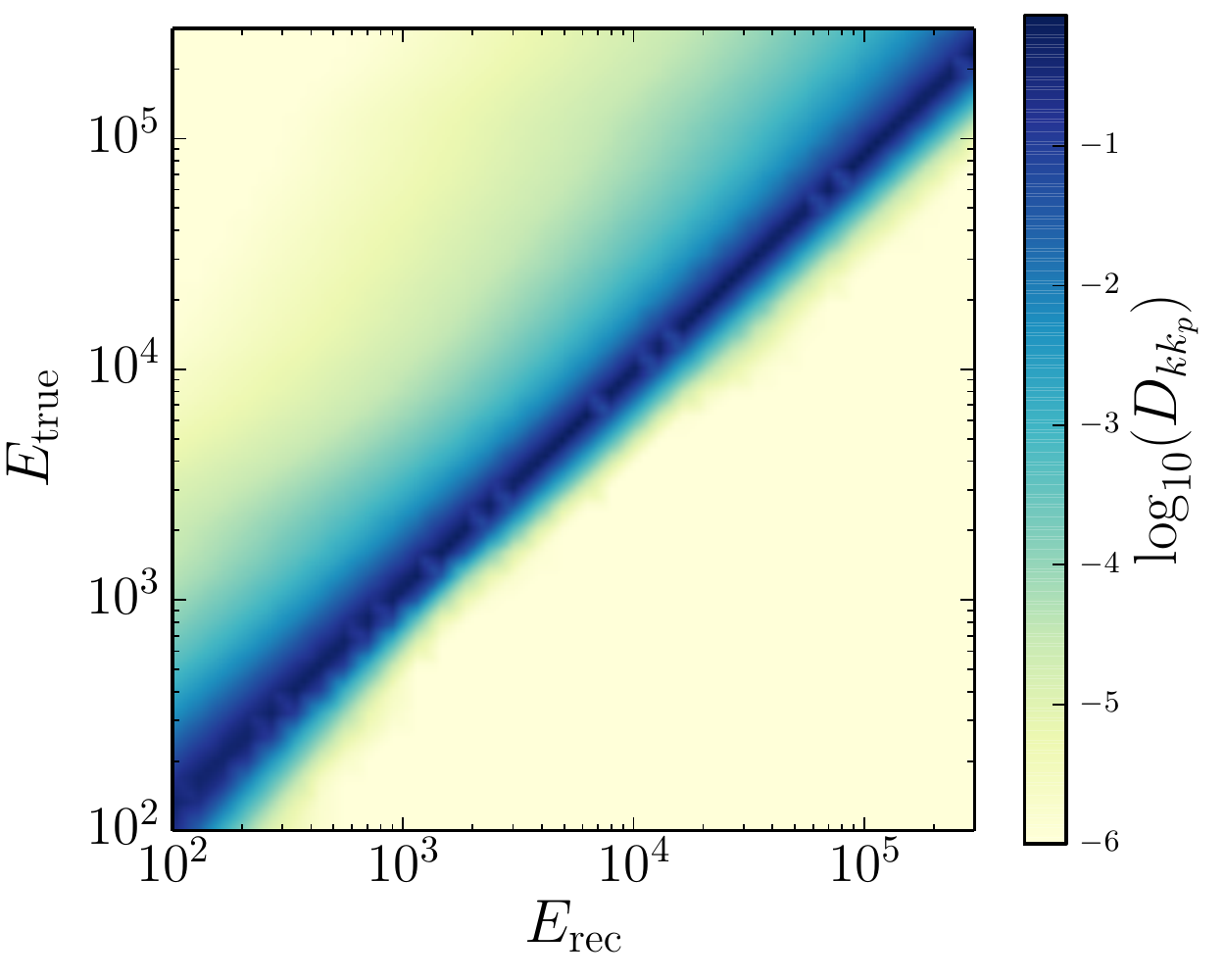}}
 \end{picture}
\end{center}
\caption{Energy dispersion matrix, $D_{kk_{p}}$, derived for all EDISP event types together. 
The color bar, i.e., $D_{kk_{p}}$, encodes the probability for a shift between reconstructed energy ($E_{\rm rec}$) and true energy ($E_{\rm true}$), here in MeV.}
\label{fig:dispersionEn}
\end{figure}

\begin {table}
\label{tab:selection}

\begin{center}
\begin{tabular}{ |c|c|c|c| } 
 \hline
 Pulsar name & \multicolumn{1}{|p{3.5cm}|}{\centering $l_{II}$ [$^\circ$]} &   \multicolumn{1}{|p{3.5cm}|} {\centering $b_{II}$ [$^\circ$] }  & \multicolumn{1}{|p{2.0cm}|}{\centering $d$ [kpc] } \\ 
 \hline
 J1420-6048 & 313.54 & 0.23 & $5.7\pm 0.9$ \\
 J1648-4611 & 339.44 & -0.79 & $4.9\pm0.7$ \\
 J1702-4128 & 344.74 & 0.12  & $4.7\pm0.6$\\
 J1718-3825 & 348.95 & −0.43 & $3.6\pm 0.4$ \\
 J2021+3651 & 75.22 & 0.11 & $10\substack{+2\\-4}$ \\
 J2240+5832 & 106.57 & -0.11 & $7.3\pm0.7 $\\ 
 \hline
 J0835-4510(Vela) & 263.552 & -2.7873 & $0.294\substack{+0.076\\-0.050}$ \\ 
 
 \hline
 
\end{tabular}

\end{center}
\caption{Selected gamma-ray pulsars (in order of right ascension) 
	used for the present analysis. The information listed includes
	Galactic longitude ($l_{II}$), latitude ($b_{II}$), as well as 
	heliocentric distance ($d$) with corresponding errors (see text for further details). 
}
\end {table}

\subsection{Data Analysis}
\label{section:data_analysis}
We use nine years of \Fermi-LAT Pass 8 data with P8R2 SOURCE V6 IRFs. The \Fermi-LAT Pass 8 data have  an  improved  angular  resolution,  a broader energy range, larger effective area, as well as reduced  uncertainties  on  the  instrumental  response  functions~\cite{ackermann_fermi_2012} compared
to previous data releases. For the determination of \Fermi-LAT source spectra, the \texttt{Enrico} scripts to calculate differential energy spectra are used \cite{sanchez_enrico_2013}. The width of the logarithmically spaced energy bins has been chosen to be 37\% of the median energy resolution. For the analysis, \textit{SOURCE} event class and  \textit{FRONT+BACK} event types has been used.
Photons with measured zenith angles greater than $90^{\circ}$ were excluded to avoid contamination by intense gamma-ray emission from the Earth's limb caused by cosmic rays interacting in the atmosphere. The region of interest (ROI) is centered on the source position and has a radius of $15^{\circ}$. We include all point sources listed in the third \Fermi-LAT source catalog~\cite{acero_fermi_2015}  within $15^{\circ}$ from the ROI center. The diffuse background is modeled with the templates for the Galactic and the isotropic extragalactic gamma-ray emission available within the \texttt{Fermi Science tools}. We keep the diffuse Galactic emission model as well as the isotropic emission model fixed for the 
flux determination in the individual energy bins after fitting it over the entire energy range.  In the spectral analysis, pulsar spectra are modeled with a power-law with  exponential cutoff:
  \begin{equation}
  \label{eqn:fitspec}
 \frac{dN}{dE}= N_{0} \left(\frac{E}{E_{0}}\right)^{-\Gamma} \exp\left(-\frac{E}{E_{\rm cut}}\right),
 \end{equation}
except for Vela, where the exponential cutoff is modified:
 \begin{equation}
 \label{eqn:velafitspec}
 \frac{dN}{dE}= N_{0} \left(\frac{E}{E_{0}}\right)^{-\Gamma_{1}} \exp\left[\left(-\frac{E}{E_{\rm cut}}\right)^{\Gamma_{2}}\right].
 \end{equation}

The free parameters are $N_{0}$ (normalization factor at the scale energy $E_{0}$),  $\Gamma$ (photon index), and $E_{\rm cut}$ (cutoff energy). For 
the Vela energy spectrum, the additional parameter $\Gamma_2$ is determined from the fit. The spectral parameters of other point sources within $3^{\circ}$ from the ROI center are left free to vary, while the parameters for the point sources at larger angles are kept fixed.  

We investigate the presence of spectral distortions due to photon-ALPs oscillations, by comparing  the goodness-of-fit with and without photon-ALPs oscillations. Similarly to a previous study to search for spectral irregularities with \Fermi-LAT \cite{ajello_search_2016}, we take into account the energy dispersion matrix 
  $D_{kk_p}$. We derive the energy dispersion matrix $D_{kk_{p}}$ via the transformation of the number of counts in true energy of a particular energy bin to the number of counts in that bin of reconstructed energy (figure~\ref{fig:dispersionEn} and see~\cite{ajello_search_2016} for further details).
  The modeled spectra are: 
  
\begin{equation}
\label{eq:fitnoALPs}
\left(\frac{dN}{dE}\right)_{\rm w/o \,  ALPs} =D_{kk_{p}} \cdot \left(\frac{dN}{dE}\right)_{\rm intrinsic} \, , 
\end{equation}
and
\begin{equation}
\label{eqn:master}
\left(\frac{dN}{dE}\right)_{\rm w \, ALPs} =D_{kk_{p}} \cdot (1- P_{\gamma \rightarrow a}\left( E,g_{a\gamma\gamma},m_{a},d\right)) \cdot \left(\frac{dN}{dE}\right)_{\rm intrinsic} \, , 
\end{equation}

where the intrinsic spectrum refers to eqs.~\ref{eqn:fitspec} and~\ref{eqn:velafitspec}.
The probability $P_{\gamma\rightarrow a}$ is calculated following the approach described in \cite{meyer_first_2013} (including the electron density model for the interstellar medium \cite{cordes_ne2001.i._2002} and a recently updated Galactic magnetic field model \cite{jansson_new_2012}).

We perform a fit to the differential flux measurements, minimizing the $\chi^{2}$ function, as has been done in previous studies ~\cite{ackermann_detection_2013,jogler_revealing_2016} and including the systematic errors estimated from the analysis of the Vela energy spectrum (see section~\ref{section:systematics}). We have verified that the 
log-likelihood as a function of flux normalization in the individual energy bins
 has indeed a parabolic shape and therefore we conclude that
  a  $\chi^{2}$-analysis for these bins are appropriate.
        
\begin{figure}
\setlength{\unitlength}{.8cm}
\begin{center}
\begin{picture}(10,12)
 \put(-4.5,0){\includegraphics[width=14cm]{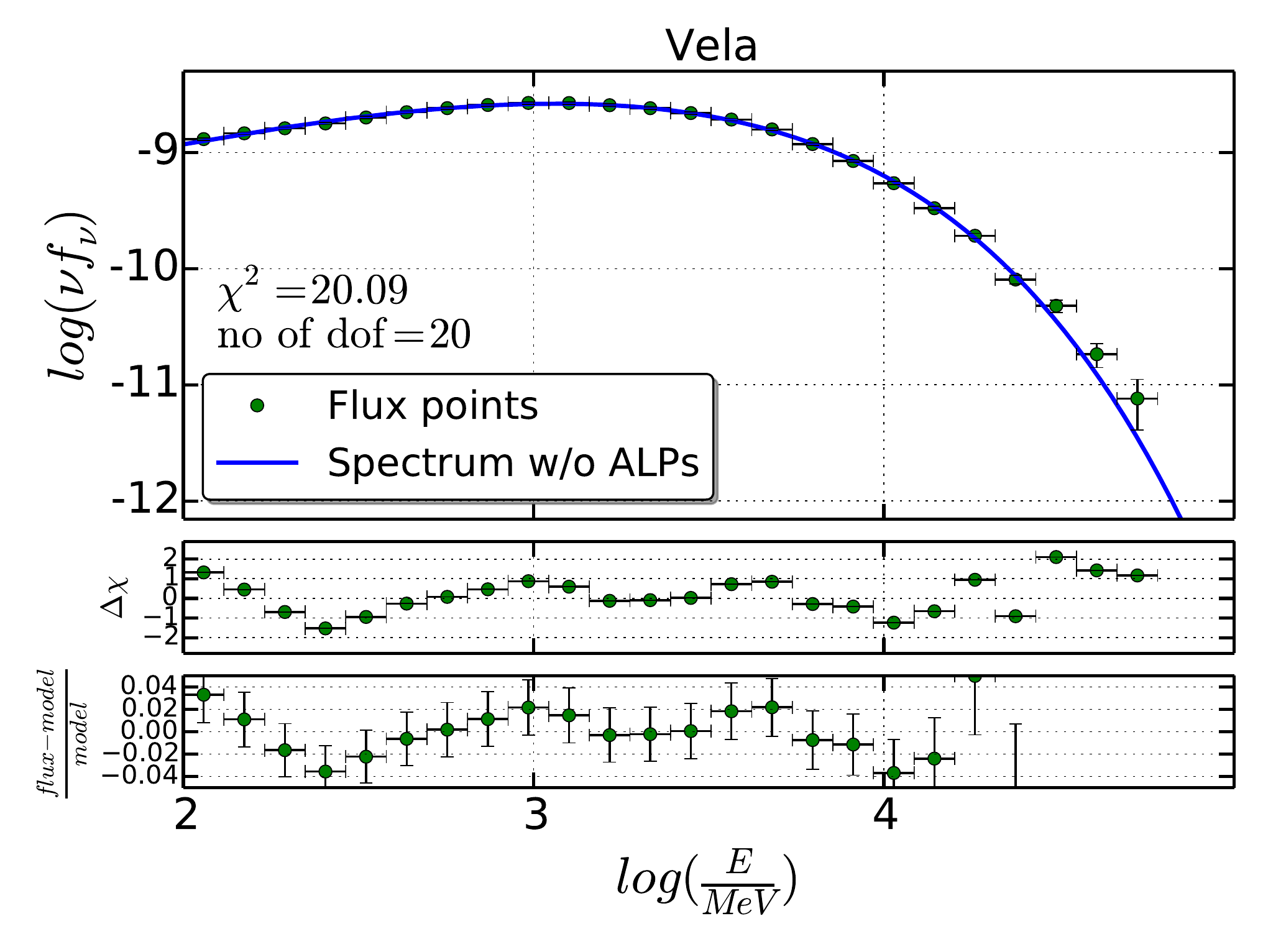}}
 \end{picture}
\end{center}
\caption{The phase-averaged energy spectrum of Vela (upper panel), 
	the residuals (middle panel), and 
	relative deviations (lower panel) overlaid with a best-fit model (eqn.~\ref{eqn:velafitspec}). Assuming a relative systematic uncertainty of the flux of 2.4~\% (added in quadrature to the statistical errors), an acceptable fit ($\chi^2(dof)=20.09(20)$) is achieved. \label{fig:espec_vela}}
\end{figure}

\begin {table}

\begin{center}
\begin{tabular}{ |c|c|c|c|c| } 
 \hline
 Pulsar name & \multicolumn{1}{|p{3cm}|}{\centering $N_0$\\ {\tiny $[10^{-9}$MeV$^{-1}$\,cm$^{-2}$\,s$^{-1}]$}} &\multicolumn{1}{|p{1.5cm}|}{\centering $E_0$ \\ {\tiny [GeV]}}&  \multicolumn{1}{|p{2cm}|}{\centering $\Gamma$} & \multicolumn{1}{|p{2.5cm}|}{\centering $E_\mathrm{cut}$ \\ {\tiny [GeV]}} \\
 \hline
 J1420-6048 & $0.0014(2)$ & 5.6 & $1.79(4)$ & $4.3(4)$ \\
 J1648-4611 & $0.0022(1)$ & 2.9 & $ 0.98(3)$ & $3.1(2)$ \\
 J1702-4128 & $0.15(3)$ &  0.1& $0.8(1)$  & $0.8(1)$  \\
 J1718-3825 & $0.021(1)$ & 1.2 & $1.58(4)$  & $2.2(2)$ \\
 J2021+3651 & $0.15(1)$ & 0.8 & $1.59(3)$ & $3.2(3)$\\
 J2240+5832 & $0.0065(1)$ &1.2 & $1.5(1)$  & $1.6(4)$\\ \hline
 J0835-4510 & 105(2)& 0.1& $\Gamma_1$ =1.27(1)&0.654(3) \\ 
 (Vela)&&&$\Gamma_2$= 0.541(2) & \\
  
 \hline
 
\end{tabular}

\end{center}
\caption{Fit results for individual pulsars without photon-ALPs mixing. The table contains the best fitted parameters  i.e., normalization factor at scale energy $(E_0)$, photon index, cutoff energy of each sources. The combined statistical and systematic (1$\sigma$) uncertainties estimated from the fit are listed as well. 
	\label{table:results_noalps}
}
\end {table}

\subsection{Systematic uncertainties}
\label{section:systematics}
In the most extensive study of the systematic uncertainties of flux measurements \cite{ackermann_fermi_2012} a number of effects contributing to systematic uncertainties are considered, including residual particle background, effective area, energy resolution, point-spread function, and  (global) uncertainties on energy scale. 
First, we discuss the effect of the uncertainties on the analysis carried out here and, secondly, we consider a robust approach to estimate the effect of uncertainties in a data-driven way.
\paragraph{Known systematics:} The effect of particle background has been checked by repeating the analysis with different event classes. For large signal-to-noise sources (as considered here), the effect is negligible.  The energy resolution has a known effect, especially at low energies and can lead to a relative bias in the reconstructed flux  by 5\% below 300~MeV for a hard spectrum (photon index $\Gamma<1.5$). The studies presented in \cite{ackermann_fermi_2012}, indicate that the resulting bias could lead to structures at around 1~GeV in the spectrum. The effect of uncertainties on the point-spread function is difficult (and in the case considered here not necessary) to distinguish from the uncertainties affecting the effective area. Finally, the energy scale calibration from beam data and in-orbit cosmic-ray data indicate that the energy scale is uncertain at the level of ($+2\%/-5\%$) in the energy range between 1~GeV--100~GeV, slightly increasing to ($+4\%/-10~\%$) below and above this energy range -- leading to a global shift of the spectral features.
\paragraph{Data driven method:} In the approach chosen here, we estimate the systematic uncertainties relevant to this analysis. Similar to the analysis carried out in \cite{ackermann_fermi_2012}, we use the energy spectrum of the Vela pulsar and derive the flux in 9 bins per decade of energy.  The resulting energy spectrum is modeled by a function of the form given in eq. \ref{eqn:velafitspec}. The parameters are estimated using a $\chi^2$-minimization which allows to quantify the goodness of fit. After inspecting the residuals, we add in quadrature to the statistical uncertainties a relative systematic uncertainty on the flux measurement. We increase the relative uncertainty until the resulting $\chi^2$ per degrees of freedom $\sim$ 1. Differently to the approach chosen in \cite{ackermann_fermi_2012}, where the envelope of flux uncertainties is considered in order to estimate the total uncertainties on the parameters estimated, we consider the minimum systematic uncertainty which leads to an acceptable fit, i.e., we increase the uncertainties such that deviations from the smooth model spectrum are not significant anymore. In this framework of determining systematics, the 
maximum relative uncertainty on the flux is 2.4\%. The result for the spectrum analysis of the Vela pulsar is presented in figure~\ref{fig:espec_vela}. Note, that at the high-energy end of the spectrum, deviations (at more than 1~$\sigma$)in excess of 4~\% of the fit appear, which are related to a power-law component in the spectrum measured with H.E.S.S. \cite{2017AIPC.1792d0028D}.
 The described method is  suitable for our purposes as we are trying to estimate the
maximum influence of systematic uncertainties leading to deviations from a smooth spectral model. This method is often applied in X-ray spectroscopy to ensure that systematic uncertainties on the flux measurement do not affect the goodness of fit anymore 
\cite{guver_systematic_2012,tsujimoto_cross-calibration_2011}.

\begin{figure}
\setlength{\unitlength}{.9cm}
\begin{center}
\begin{picture}(12,7)
 \put(-3.2,0){\includegraphics[width=8.6cm]{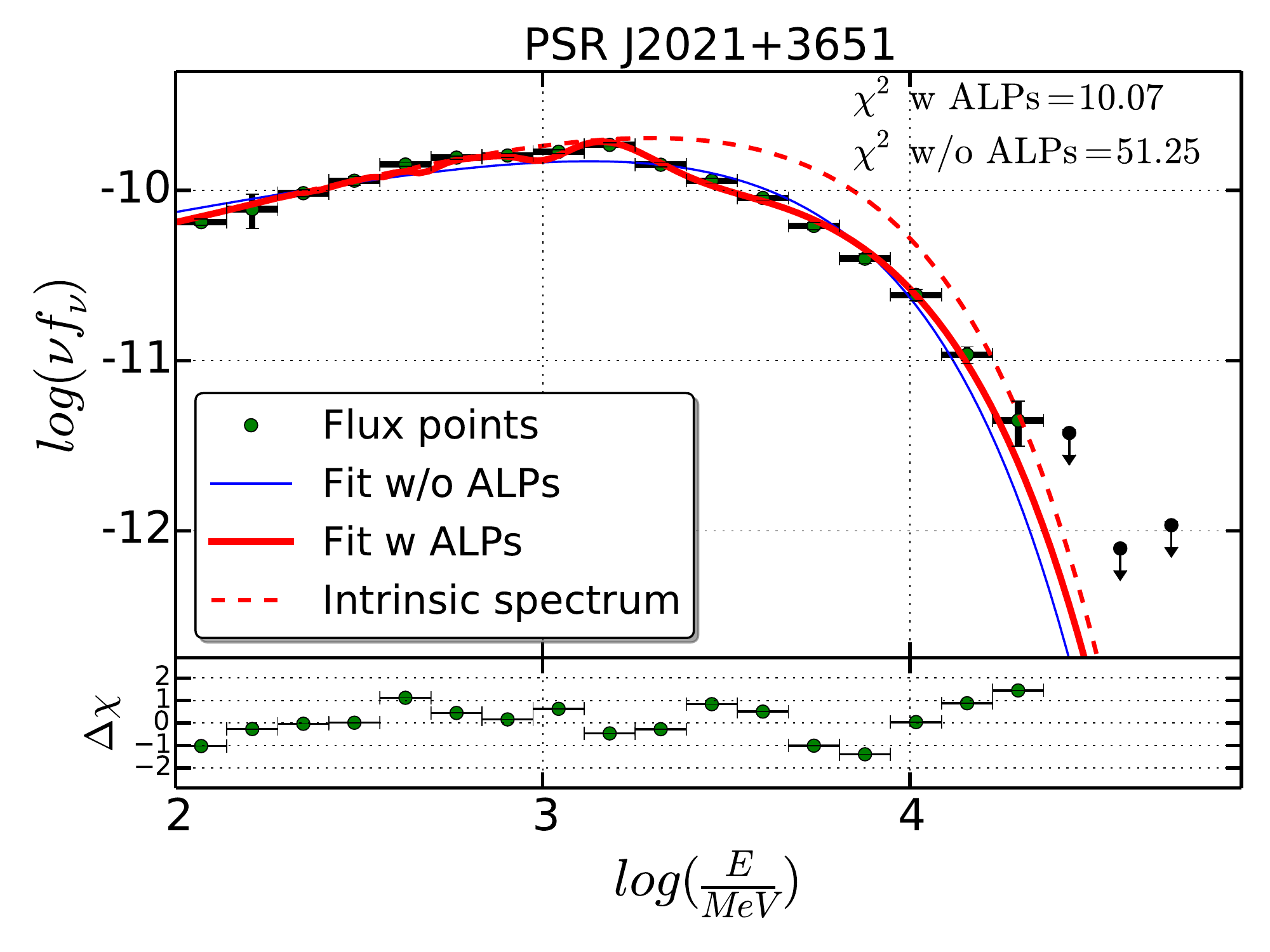}}
 \put(6.2,0){\includegraphics[width=8.6cm]{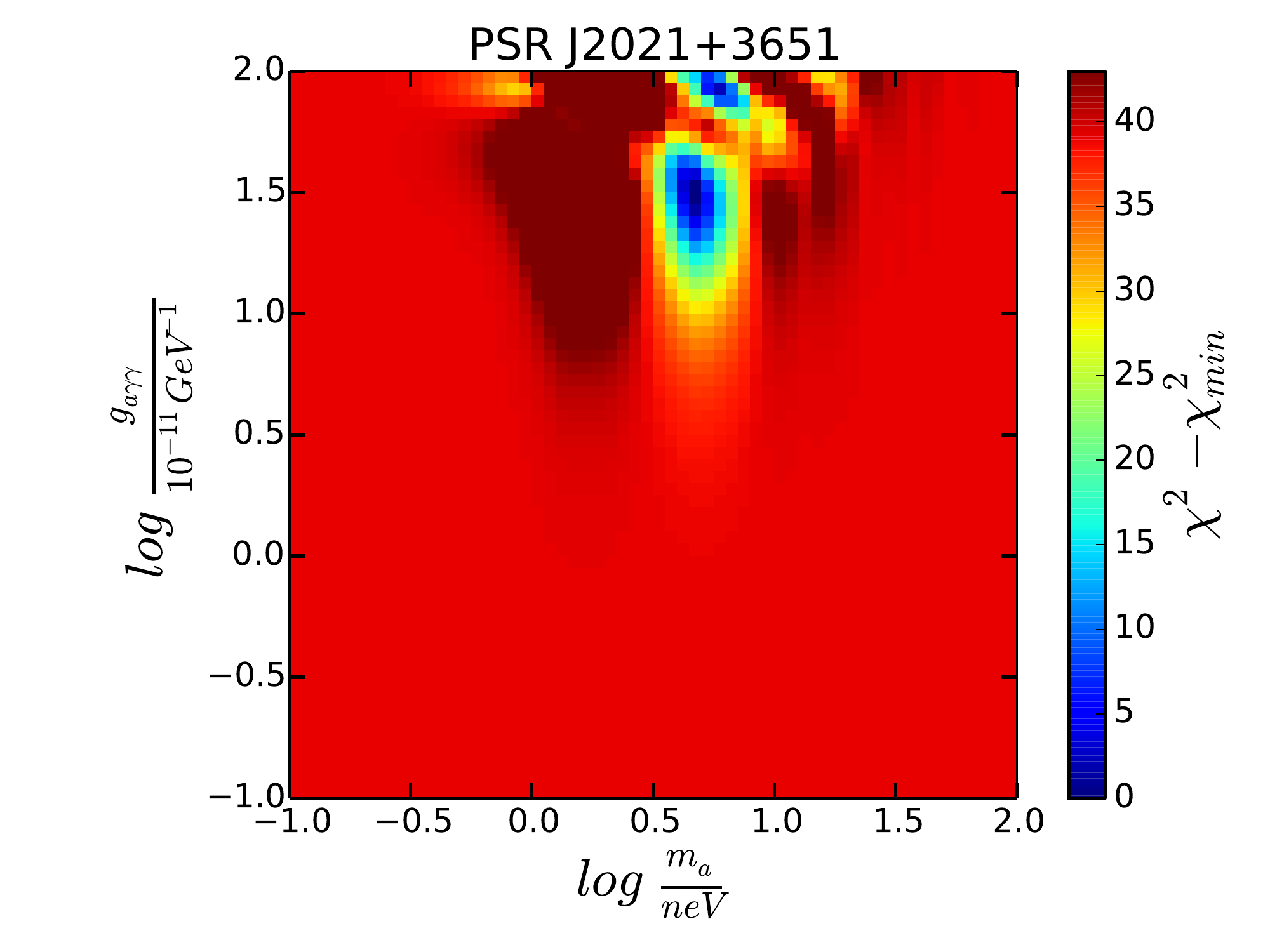}}
 \end{picture}
\end{center}
\caption{\label{fig:j2021}
	Left panel: The spectral energy distribution of PSR~J2021+3651 
	(green points with combined systematic and statistical uncertainties) overlaid with the best-fit
	models (blue thin: without photon-ALPs mixing, red thick: with photon-ALPs mixing).
	In addition, the intrinsic spectrum (as emitted) is shown by the red dashed line to highlight the effect of photon-ALPs mixing.
	Right panel: A scan of the plane of mass and coupling, where the color scale indicates
	the increase of $\chi^2$ with respect to the global minimum. }
\end{figure}
\begin{small}
\begin {table}

\begin{center}
\begin{tabular}{ |c|c|c|c|c|c|c| } 
 \hline
 Pulsar name & \multicolumn{1}{|p{3cm}|}{\centering $N_0$\\ {\tiny $[10^{-9}$MeV$^{-1}$\,cm$^{-2}$\,s$^{-1}]$}}&
   \multicolumn{1}{|p{2cm}|}{\centering $\Gamma$} & \multicolumn{1}{|p{1.5cm}|}{\centering $E_\mathrm{cut}$ \\ {\tiny [GeV]}}&
  \multicolumn{1}{|p{2.3cm}|}{\centering $g_{a\gamma\gamma}$ \\
  	{\tiny $\left[10^{-10}\,\mathrm{GeV}^{-1}\right]$}} & \multicolumn{1}{|p{1.8cm}|}{\centering $m_a$\\
  		{\tiny $[\mathrm{neV}]$}}  \\
 \hline
 J1420-6048 & $0.0016(2)$ & $ 1.74(4)$ & $ 5.4(6)$& $ 1.7(3)$ & $ 3.6(1)$ \\
 J1648-4611 & $0.0028(2)$ & $ 0.88(3)$ & $ 3.4(2)$ & $ 5.3(9)$ & $ 4.3(1)$ \\
 J1702-4128 & $0.13(3)$ & $ 0.9(1)$ & $ 1.0(2)$ & $ 4.4(2)$ & $ 8.1(5)$  \\
 J1718-3825 & $0.024(2)$ & $ 1.48(4)$ & $ 2.1(1)$ & $ 2.4(3)$ & $ 8.9(2)$ \\
 J2021+3651 & $0.18(1)$ & $ 1.45(3)$ & $ 3.5(1)$& $ 3.5(3)$ & $ 4.4(1)$ \\
 J2240+5832 & $0.005(1)$ & $ 1.5(1)$ & $ 2.4(6)$& $ 2.1(4)$ & $ 3.7(3)$  \\

  
 \hline
 
\end{tabular}

\end{center}
\caption{Fit results for individual pulsars with photon-ALPs mixing. The table gives the best fitted parameters  i.e., normalization factor of each source defined at scale energy ($E_0$, see table~\ref{table:results_noalps}), 
	spectral index, cutoff energy, photon ALPs coupling constant ($g_{a\gamma\gamma}$), and ALPs mass ($m_{a}$) of each source including uncertainties. 
	\label{table:fitresults_alps}}
\end {table}

\end{small}
\section{Results}
\subsection{Energy spectra and fits}
\label{section:results}
The results of the spectral analysis and fitting of exponential power-law
models to the spectral points, i.e., fit without photon-ALPs mixing, are summarized in table~\ref{table:results_noalps}. There, we list for each pulsar
the best-fit normalization factor $N_0$, the photon index $\Gamma$, and the cutoff energy $E_\mathrm{cut}$.  In figure~\ref{fig:j2021}, the spectral energy distribution for one particular source (PSR~J2021+3651) is shown, 
overlaid with the best-fitting
model from eq.~\ref{eqn:fitspec} (blue thin line). Obviously the resulting 
$\chi^2=51.25$ with $14$ degrees of freedom for that source is not satisfactory (see table~\ref{table:chi2}, second column for the resulting $\chi^2$-values 
for all the considered pulsars). 

We consider as an alternative hypothesis, that the observed energy spectra are
modified by photon-ALPs mixing in the intervening Galactic magnetic fields (see eq.~\ref{eqn:master}). 
 
 Including the effect of spectral modulation from photon-ALPs oscillations
 improves the goodness of fit consistently  for the selected sources (see table~\ref{table:fitresults_alps} for the best-fitting values
 and third column for the resulting $\chi^2$-values). We discuss 
 in the following the most significant source PSR J2021+3651 before including the other sources in a combined analysis (figure~\ref{fig:j2021}). 
 
 With the introduction of two
 additional free parameters ($g_{a\gamma\gamma}$ and $m_a$) we can re-fit the
 spectrum of PSR~J2021+3651 and achieve with  $\Delta\chi^2=41.2$. Upon closer inspection of the energy spectrum (figure~\ref{fig:j2021} left panel), 
 the improvement is a result of the apparent 
 deviation of a smooth power-law at an energy of about 2 GeV and a flux dropping off at higher energies modifying an exponential cutoff. Both features are 
 well-described by the characteristic modulation of the photon-ALPs oscillation. We note that the shape of the modulation is directly linked to the strength of the 
 transversal magnetic field and extension of the spiral arms traversed by the line of sight.
 
 The effects of mixing  are illustrated in figure~\ref{fig:los}, 
 where the intensity of an unpolarized
 photon beam at energy 3 GeV and distance 10 kpc in the direction of PSR~J2021+3651 is followed through the magnetic field of the intervening
 interstellar medium. 
 For the favored coupling and mass, the photon intensity is reduced to roughly 60\% of the initial value. The oscillation length is 
 similar to the distance leading to a noticeable increase as well as a decrease of
 photon intensity along the line of sight.
 
 When scanning the parameters of mass and coupling (figure~\ref{fig:j2021}~right panel), there
 are quite narrow minima in the plane of $\chi^2$ which are aligned along the
 direction of larger coupling. Turning back to figure~\ref{fig:los},
  this repetitive pattern
 is the result of multiple oscillations for larger values of the coupling
  along the line of sight. For decreasing coupling, 
 the case of no-mixing is recovered. The local minima are adjacent to local maxima
 which lead to a tight constraint on the mass parameter.

  Similar improvement to the goodness of fit ($\Delta \chi^2$) can be seen 
 for the other five objects considered. The resulting best-fit parameters (including
 the re-fit spectral parameters) are listed in table~\ref{table:fitresults_alps}. The favored mass range is
 similar among the objects to be around 3~neV with a coupling between 1.7 and
 5.3 (in units of $10^{-10}$~GeV$^{-1}$). The improvement in $\chi^2$, the 
 resulting degrees of freedom for the individual spectra are listed in table~\ref{table:chi2}.
  
 The observed energy spectra and best-fit models for the other objects are shown in the appendix (figures~\ref{fig:J1420_J1648} and \ref{fig:J1702_J2240}). While for all spectra similar improvements are seen, there is an indication that  the modulation in the spectra are very similar for  objects which are aligned in the same region of the Galaxy (e.g. PSR~J2021+3651 and PSR~J2240+5832, similarly the pair PSR J1702-4128 and PSR J1718-3825) (see section~\ref{sec:summary} for a discussion of this observation).

\begin {table}

\begin{center}
\begin{tabular}{ |c|c|c|c|c| } 
 \hline
 Pulsar name  &   
 \multicolumn{1}{|p{2.5cm}|}{\centering $\chi^2(dof)$ $H_0$} &
 \multicolumn{1}{|p{2.5cm}|}{\centering $\chi^2(dof)$ $H_1$}  &
 \multicolumn{1}{|p{2.5cm}|}{\centering{Significance ($H_1/H_0$)}} & 
 \multicolumn{1}{|p{2.5cm}|}{\centering $\chi^{2}(dof)$ $H_2$}\\ 
 \hline
 J1420-6048 & 31.10(15) &21.27(13) & $1.38~\sigma$& 22.46(15)  \\
 J1648-4611 & 47.15(14) &21.37(12) & $2.38~\sigma$& 41.61(14)  \\
 J1702-4128 & 12.70(8)  &3.57(6)   & $2.01~\sigma$& 8.54(8)  \\
 J1718-3825 & 53.57(15) &25.61(13) & $2.40~\sigma$& 29.52(15)   \\
 J2021+3651 & 51.25(14) &10.07(12) & $3.86~\sigma$& 41.85(14)   \\
 J2240+5832 & 19.66(11) &8.01(9)   & $2.11~\sigma$&  8.39(11)   \\ \hline
 Combined   & 215.42(77) &89.9(65)  & $5.52~\sigma$&152.37(75)  \\
 
  
 \hline
 
\end{tabular}

\end{center}
\caption{ A comparison of the $\chi^2$ values obtained for the three hypotheses: $H_0$: no ALPs oscillation, $H_1$: ALPs oscillation with values of coupling and mass left free for individual sources, $H_2$: ALPs oscillation for a global estimate of coupling and mass. The significance is calculated using the excess variance technique (see section~\ref{section:significance} for further details.
\label{table:chi2}}
\end {table}

\subsection{Significance level}
\label{section:significance}
In order to compute the significance level in table~\ref{table:chi2}, we use the excess variance technique which is based upon the F-test for the two hypotheses: $H_0$, i.e.~no-ALPs, see eq.~\ref{eq:fitnoALPs} and $H_1$, i.e.~photon-ALPs mixing included, see eq.~\ref{eqn:master}. 
Assuming a sample size $n$, $k$ and $m$ parameters for hypotheses $H_0$ and $H_1$ respectively,   we construct the following quantity:
\begin{equation}
\label{eq:ftest}
f:=\frac{(\chi^2_{\rm H_{0}}-\chi^2_{\rm H_{1}})/(m-k)}{\chi^2_{\rm H_{1}}/(n-m)}\sim F_{m-k,n-m}.
\end{equation}
 The quantity is distributed as
the $F$-distribution with $m-k$ degrees of freedom for the summed squares in the nominator
and $n-m$ degrees of freedom in the denominator. 
The significance of the result has been estimated to be 5.52$\sigma$ for the combined sample ($H_1$). We list the
corresponding values for  the other pulsars  as well as for the combined data in table~\ref{table:chi2}. 
We also consider the hypothesis $H_2$, where we carry out a $\chi^2$-minimization of all spectra with 
a common value of $g_{a\gamma\gamma}$ and $m_a$. For this case, the overall fit deteriorates and the
resulting $\chi^2$ value is shown in table~\ref{table:chi2}, the significance for this hypothesis
is $4.6\sigma$ using eq.~\ref{eq:ftest} (see also discussion in section~\ref{sec:summary}).

\subsection{Combined fit and parameter estimate}
\begin{figure}
\setlength{\unitlength}{.9cm}
\begin{center}
\begin{picture}(10,12)
 \put(-5,0){\includegraphics[width=15cm]{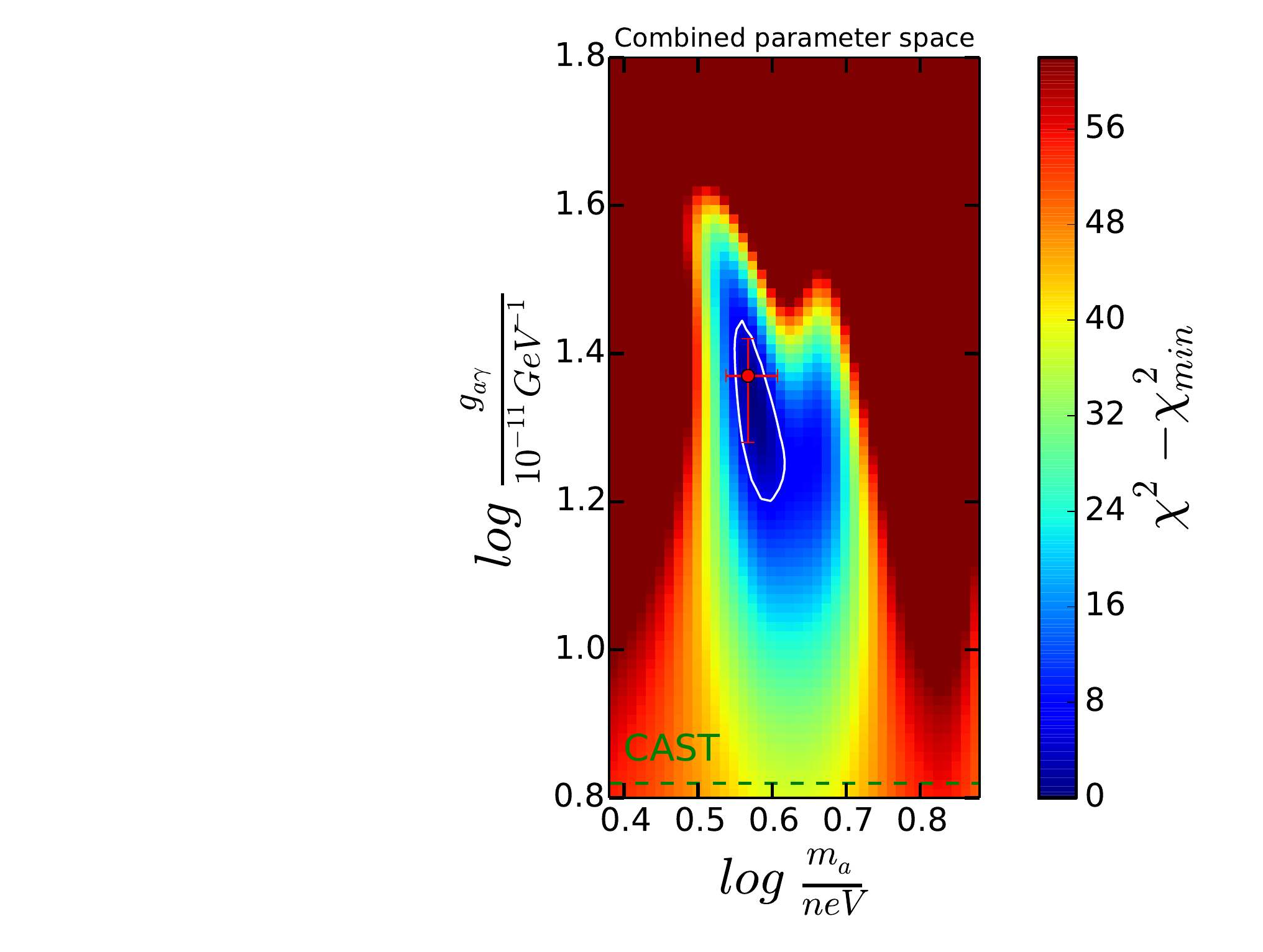}}
  \end{picture}
\end{center}
\caption{Significance map of combined  $ \chi^{2}$  analysis for the pulsars. The white marked region in the ($g_{a\gamma\gamma}, m_{a}$) plane indicates the photon-ALPs mixing contour with 95\% confidence level. The red point with the uncertainty refers the minimum position in the ALPs parameter space and projected uncertainties with 68\% confidence level. Green horizontal line represents the upper limit on the photon-ALPs coupling strength $g_{a\gamma\gamma}$ of the CERN Axion Solar Telescope (CAST)~\cite{cast_collaboration_new_2017}. 
 \label{fig:combined}}

\end{figure}

After we have established that the ALPs-hypothesis provides a significantly
better description of the data, we continue and estimate the best-fitting
ALPs-related parameters
(mass and coupling) by summing the individual $\Delta\chi^2$ planes of the six
source spectra. The result is shown in figure~\ref{fig:combined}.
We find for the best estimate is the coupling $g_{a\gamma\gamma}=(2.3\substack{+0.3\\-0.4}) \times 10^{-10} \rm GeV^{-1} $ and ALPs mass $m_a=(3.6\substack{+0.5\\-0.2})$ neV. The $2\,\sigma$-uncertainty contour is marked by the white line in the same figure.

We estimate the systematic uncertainties related to the magnetic field
strength and the uncertainties of the distance:
We modify the magnetic field within the quoted uncertainties of the respective model, in order to understand the effect of its variation on our mixing contours. With an increase of 20\% of the magnetic field along the line of sight, the
 coupling constant is reduced by 20\% changing from $3.5 \times 10^{-10} \, \rm GeV^{-1}$  to $2.8 \times 10^{-10}  \, \rm GeV^{-1}$ (see figure~\ref{fig:Bfield_uncer_spectrum}). Similarly, 40\% enhancement in the magnetic field intensity brings the coupling constant even lower. In both  cases, $ \chi^{2}$ decreases slightly which implies that the overall fit favors an increased Galactic magnetic field.

 For PSR~J2021+3651, the effect of the distance uncertainty is most pronounced. Given the rather large uncertainty on the distance, the
 object is located either in front of or even behind the fifth
 spiral arm. Reducing the distance by 4 kpc, we obtain a change $\approx 2.4\times 10^{-10} \, \rm GeV^{-1}$, corresponding to around 70\% enhancement in $g_{a\gamma\gamma}$, while the ALPs mass increases by 0.86~neV. When increasing the distance by 2 kpc, instead, $g_{a\gamma\gamma}$ changes by 24\%, i.e.,~$g_{a\gamma\gamma} \sim 2.7 \times 10^{-10} \, \rm GeV^{-1}$ and the mass varies around 1 neV. The 
 corresponding spectral fits associated with this analysis are shown in figure~\ref{fig:distance_uncer_spectrum} in the appendix.
  In order to estimate the uncertainties related to the estimate of the global
  parameters for mass and coupling, we increase the magnetic field by 20~\% for all sources and increase the distance within the uncertainties. The resulting
  best-fit values are used to estimate the systematic uncertainties to be for mass $m_a=(3.6\substack{+0.5_ \mathrm{stat.}\\-0.2_ \mathrm{stat.}}\pm 0.2_\mathrm{syst.} )$~neV and $g_{a\gamma\gamma}=(2.3\substack{+0.3_ \mathrm{stat.}\\-0.4_ \mathrm{stat.}}\pm 0.4_\mathrm{syst.})\times 10^{-10}$~GeV$^{-1}$.

\section{Summary and conclusion}
\label{sec:summary}

In this article, we study for the first time modulations in the gamma-ray spectra of bright Galactic pulsars induced by photon-ALPs mixing in the Galactic magnetic field.
With the \Fermi-LAT dataset of nine years from six different pulsar candidates selected
according to their location in the Galaxy and brightness,  we investigate the presence of the spectral irregularities. 
We find  evidence (at the $5.52\sigma$-level) 
for the presence of spectral irregularities, absent in the nearby bright Vela pulsar.  While the spectral variations are as large as $20\,\%-40\,\%$, the
maximum systematic relative flux uncertainties found for the Vela spectrum 
is $2.4~\%$.\\
In the combined analysis, we estimate  $g_{a\gamma\gamma}=(2.3\substack{+0.3_ \mathrm{stat.}\\-0.4_ \mathrm{stat.}}\pm 0.4_\mathrm{syst.})\times 10^{-10}$~GeV$^{-1}$  and 
$m_a=(3.6\substack{+0.5_ \mathrm{stat.}\\-0.2_ \mathrm{stat.}}\pm 0.2_\mathrm{syst.} )$ neV.
We note, that the combined data-set is not well-described by a fixed value of
photon-ALPs coupling and mass (table~\ref{table:chi2}, marked as $H_2$). The differences of the mass, coupling for individual lines of sight (table~\ref{table:fitresults_alps}) are similar but not consistent within the
statistical uncertainties.  Mass and coupling should be unified for all lines of sight. However, we do have limited knowledge about the magnetic field structure, especially for the sources which are located in the inner part of the Galaxy (see figure~\ref{fig:los}) - we also note that the crucial opening angle of the spiral arms is not well constrained. The magnetic field models are derived on the basis of Faraday-rotation measures which are sensitive only to the
longitudinal magnetic field which is not of relevance for photon-ALPS coupling.
 Additionally, in the inner Galaxy the structure of spiral arms is not well resolved and unknown magnetic field components could be present. We note that the good fit of the model with slightly varying values of mass and coupling does indeed produce an acceptable fit (hypothesis $H_1$ marked in table~\ref{table:chi2}). 
\begin{figure}
\setlength{\unitlength}{.9cm}
\begin{center}
\begin{picture}(10,10)
 \put(-3,0){\includegraphics[width=14cm]{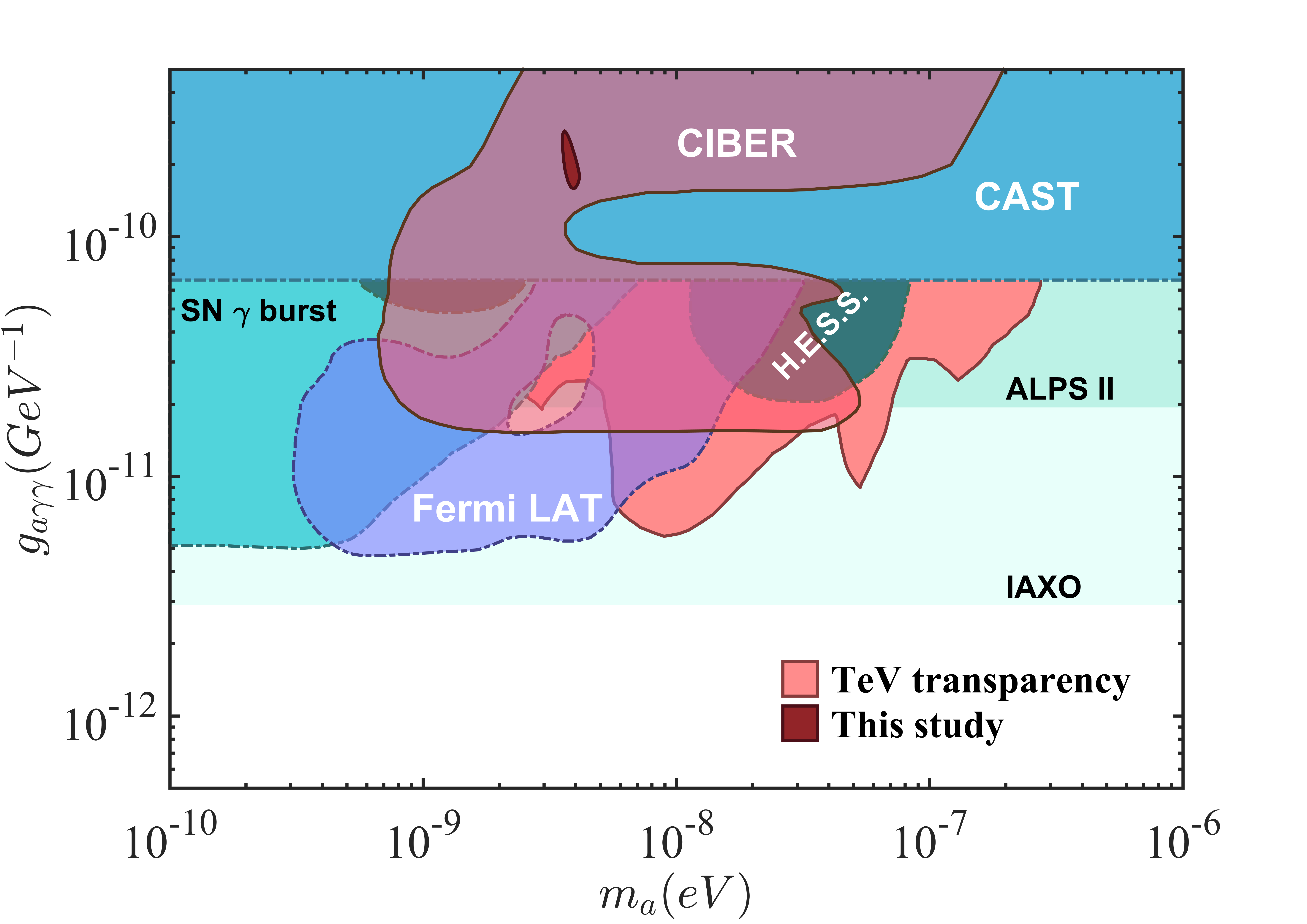}}
  \end{picture}
\end{center}
\caption{Limits on ALPs parameter space in the ($m_{a}, g_{a\gamma\gamma}$) plane. The parameter space surrounded by solid lines present the hints from ALPs. The horizontal light sky blue bands are shown as the sensitivity of ALPS-II and IAXO experiments. The regions enclosed by dotted lines and different shades in blue represent the constraints on ALPs contour given by different observations and experiments. The brown-shaded contour represents the 
	parameters estimated from pulsar spectra as found by the present analysis.
	\label{fig:masterplot}  }
\end{figure}

 The favored 2$\sigma$ contour derived from this analysis is compared with the other existing results in figure~\ref{fig:masterplot}. The best-fit parameters are well consistent
 with the lower-limit analysis related to the TeV transparency \cite{meyer_first_2013} as well
 as the a similar analysis marked CIBER \cite{kohri_axion-like_2017}. There is  no obvious
 conflict with the constraints derived from searches for irregularities in gamma-ray spectra from PKS~2155
 (HESS \cite{abramowski_constraints_2013}) and NGC1275 (\Fermi-LAT \cite{ajello_search_2016}). 
 At a first glance, the non-observation of a
 prompt gamma-ray signal from SN1987 \cite{payez_revisiting_2015} and the limit from the CAST helioscope \cite{cast_collaboration_new_2017} are in tension with the signal observed here. It is however important to note that the conversion of photons into ALPs is presumed to take place in an environment (inside 
 a star) distinctly different from the dilute interstellar medium where the conversion is occurring in the case here analyzed.
 
We also note that the signal is well within reach of the upcoming ALPS-II 
 light shining through wall experiment \cite{bahre_any_2013}.
 
 Since the objects observed are pulsars, there may be a source intrinsic effect (even though the Vela pulsar does not show any modulations). In a recent study of 
 the extended Galactic supernova remnant IC433, a similar type of analysis was carried
 out with consistent results \cite{2018arXiv180101646X} which strengthens the case for an explanation
 which is not related to the source or its emission process. 
 
 At present, we have not been able
 to identify a known propagation effect which could lead to a similar type of spectral
 modulation.

\acknowledgments
We acknowledge fruitful discussions with M.~Meyer, A.~Mirizzi and T.~R.~Slatyer. 
In particular, we thank M. Meyer for providing us with the script to compute the energy 
dispersion matrix.  JM would like to thank for the financial support through the 
DFG funded collaborative research center SFB 676 \textbf{Particles, Strings,
and the Early Universe}. 
FC acknowledgdes the hospitality of the SFB 676. 
This research has been done with the use of \texttt{Fermi Science Tools} 
developed by the \Fermi-LAT collaboration.

\label{Bibliography}

\bibliographystyle{ieeetr}

\bibliography{axionrelated}

\section{Appendix}

\subsection{Pulsar spectra}

\begin{figure}
\setlength{\unitlength}{.9cm}
\begin{center}
\begin{picture}(12,16)
 \put(-3.2,0){\includegraphics[width=8.6cm]{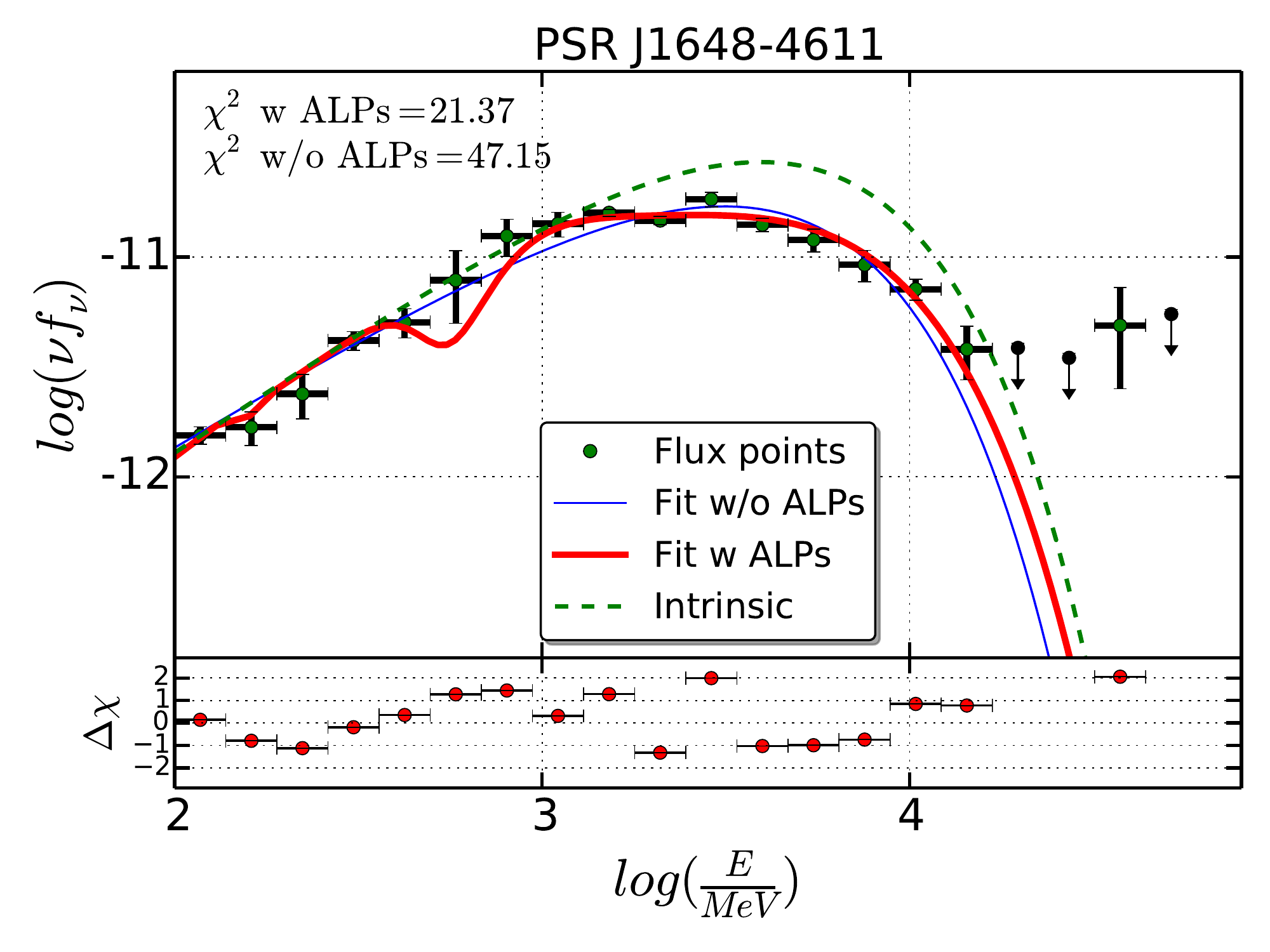}}
 \put(6.2,0){\includegraphics[width=8.6cm]{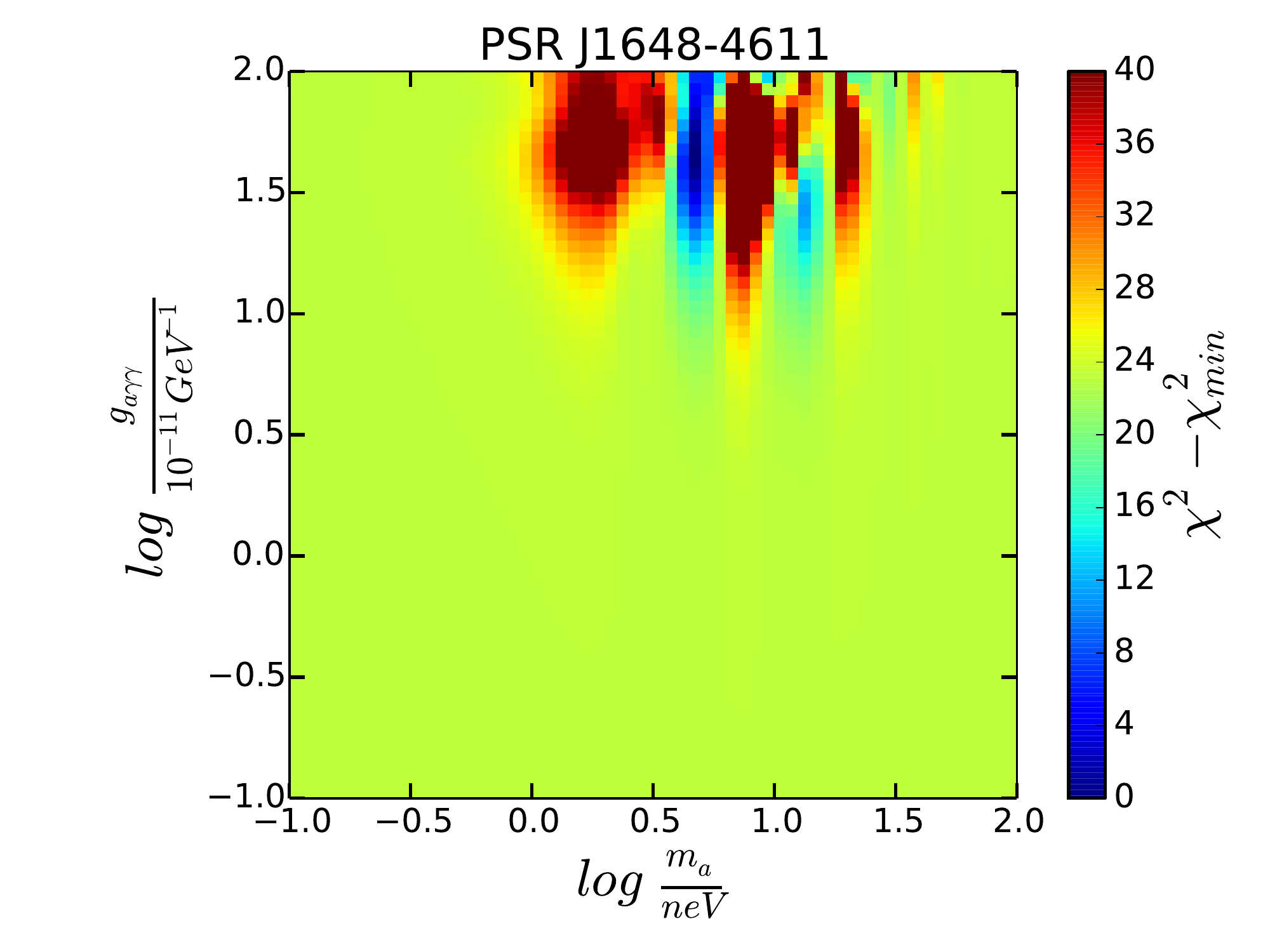}}
 \put(-3.2,8){\includegraphics[width=8.6cm]{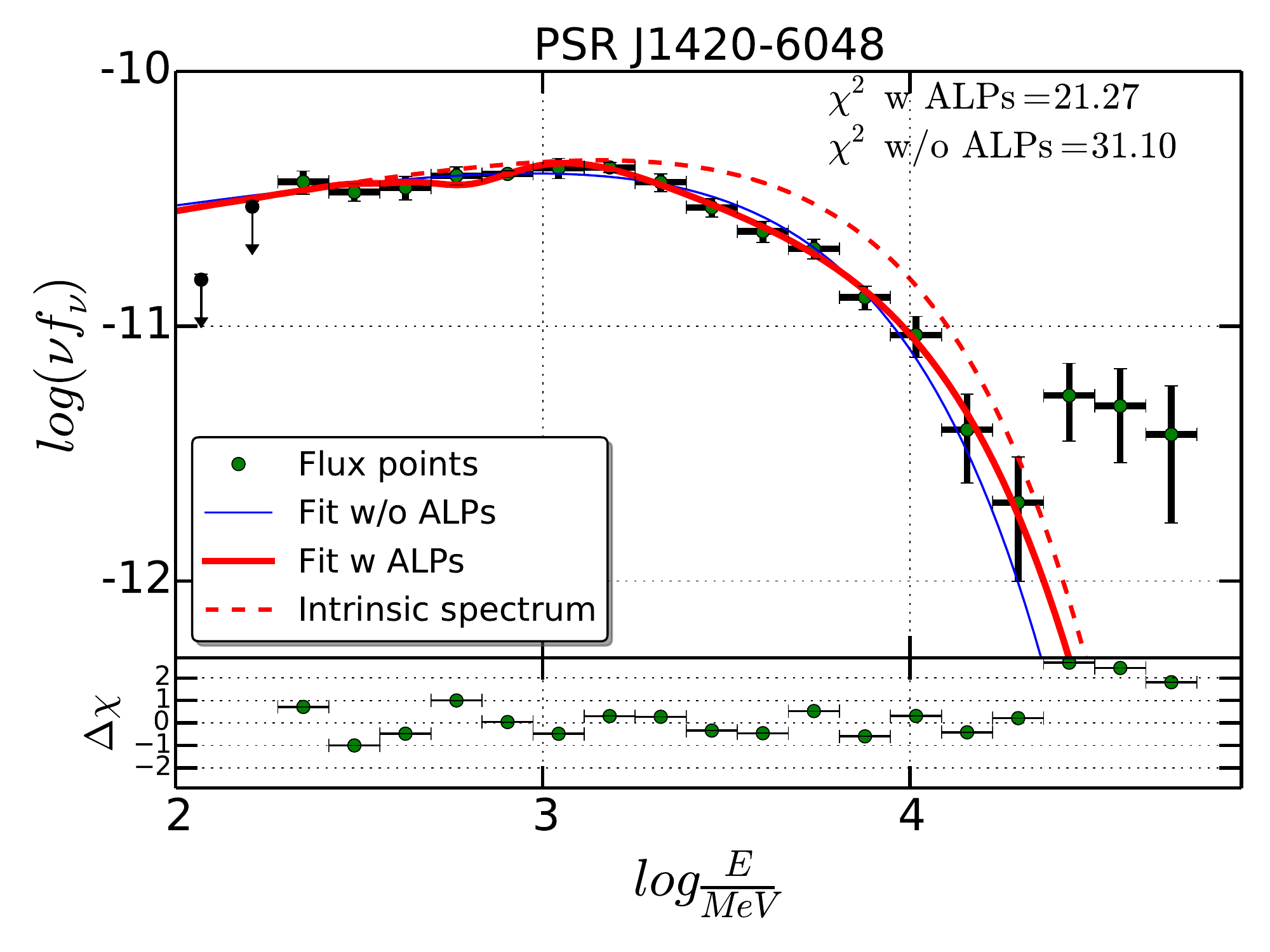}}
 \put(6.2,8){\includegraphics[width=8.6cm]{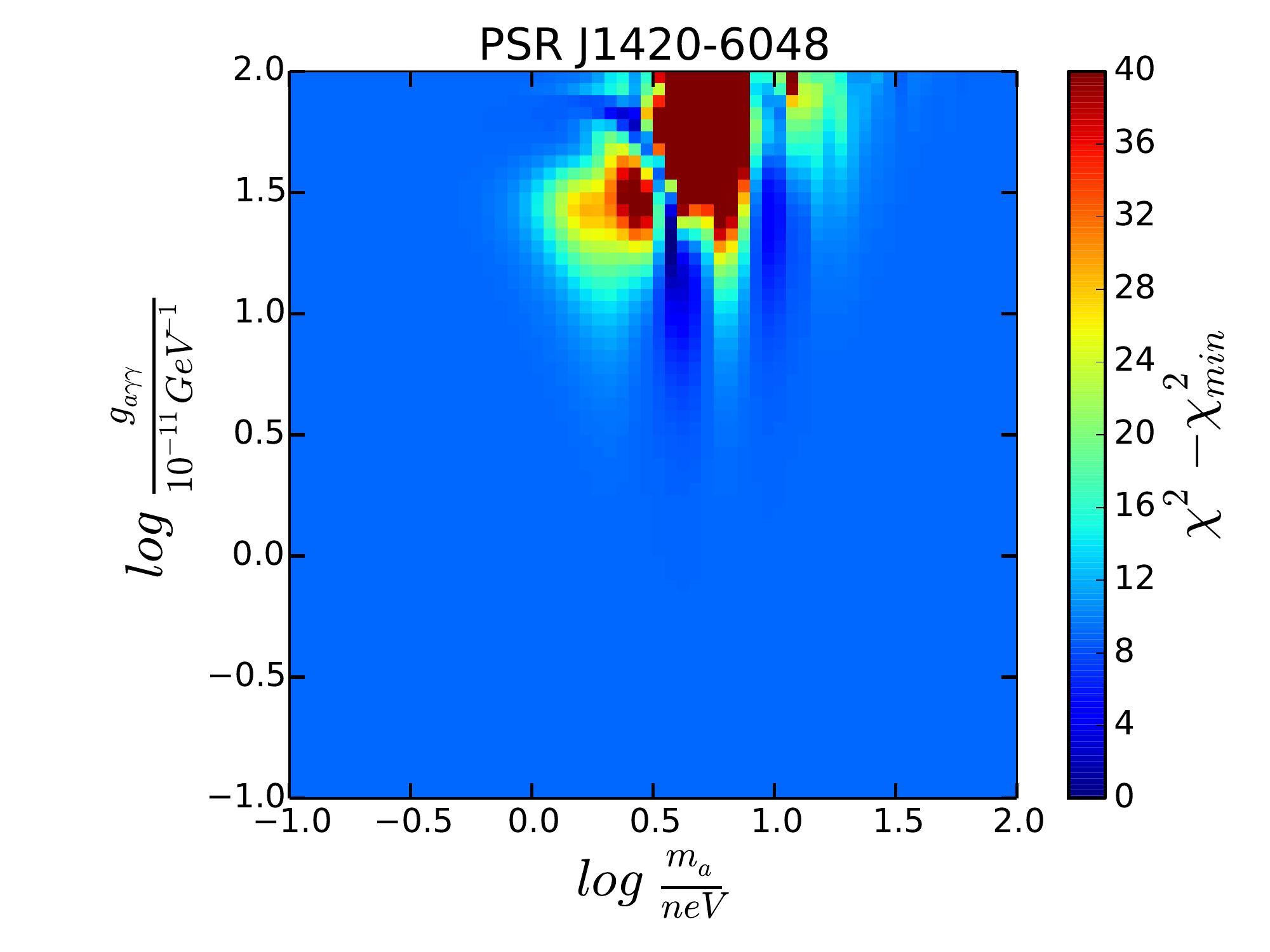}}
 \end{picture}
\end{center}
\caption{Upper panel:  Spectrum and best-fit contour plot of PSR J1420-6048. Lower panel: Spectrum and best-fit contour plot of PSR J1648-4611. In the left column, the green points correspond to the energy flux points derived using Fermi-LAT binned analysis, blue line refers to the best fit model to the flux points without ALPs parameters, whereas the red line follows the best fit model to the flux points with ALPs parameters. $\triangle\chi$ values has been plotted in the bottom panel of each spectrum plot. In the right column, the best fit contour has been illustrated in the($g_{a\gamma\gamma}, m_{a}$) plane where, the lower values in the colorbar gives the best fit mixing region. 
\label{fig:J1420_J1648} }

\end{figure}

\begin{figure}
\setlength{\unitlength}{.9cm}
\begin{center}
\begin{picture}(12,19)
\put(-3.2,15){\includegraphics[width=8.6cm]{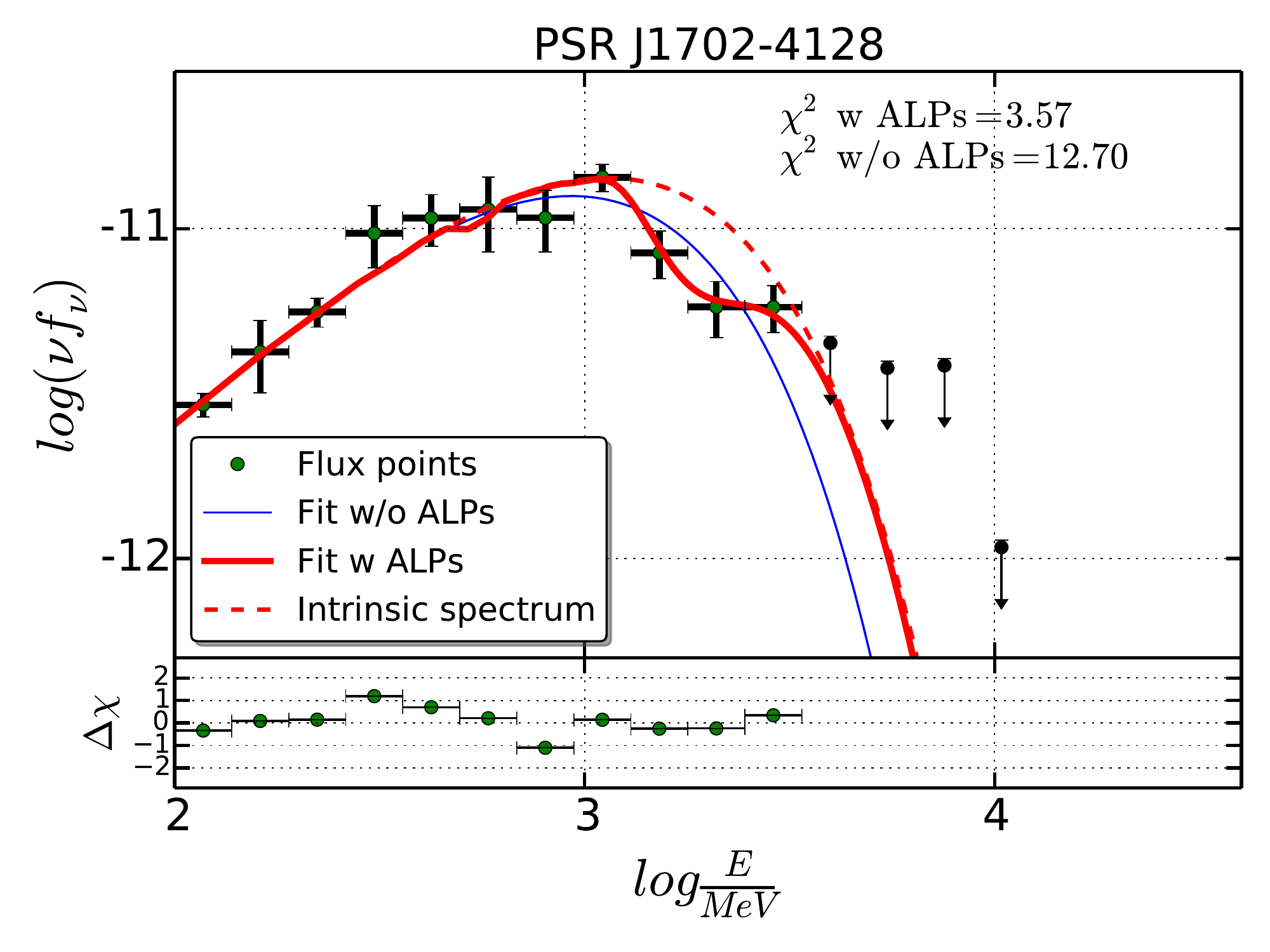}}
 \put(6.2,15){\includegraphics[width=8.6cm]{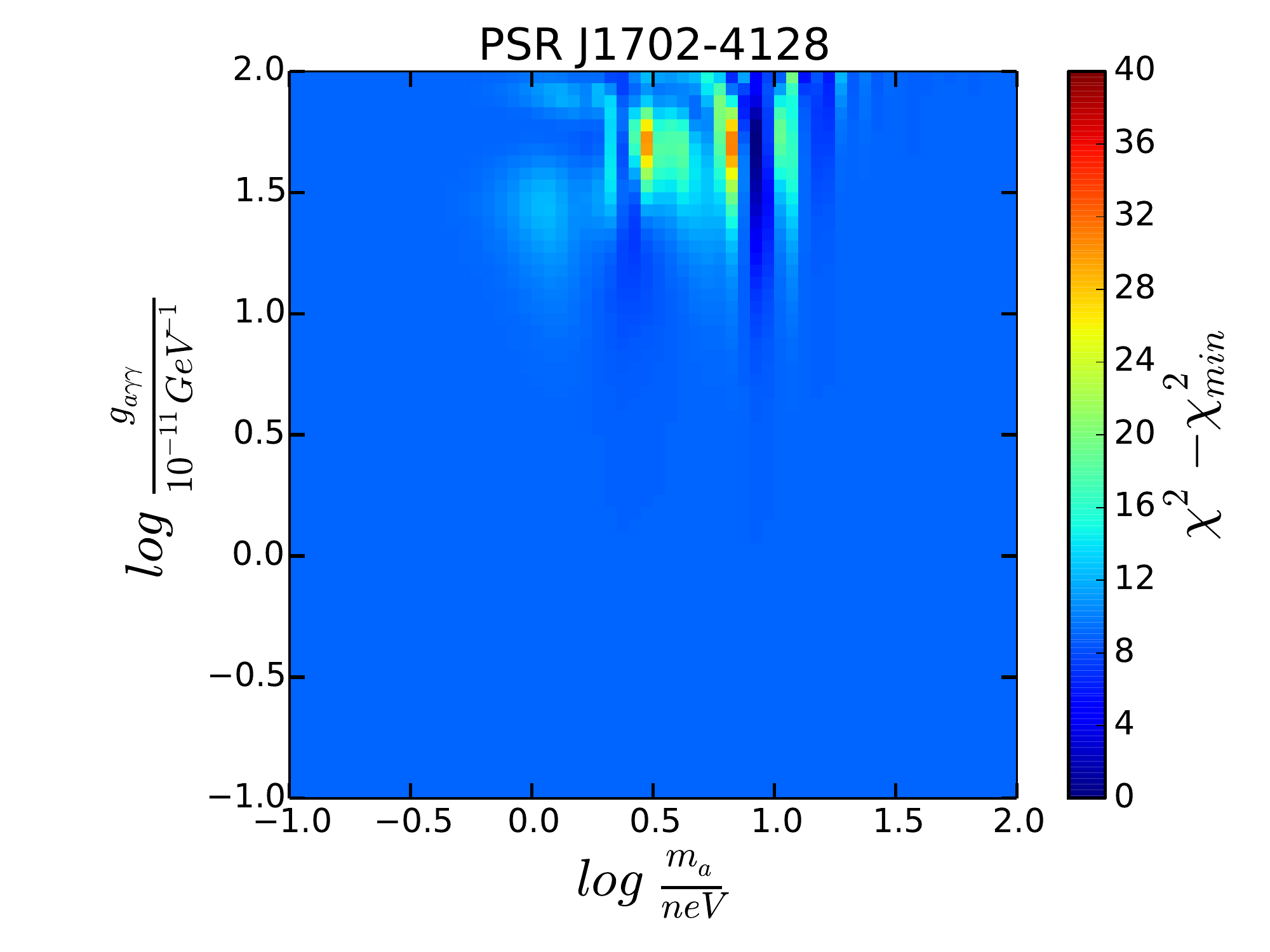}}
 \put(-3.2,0){\includegraphics[width=8.6cm]{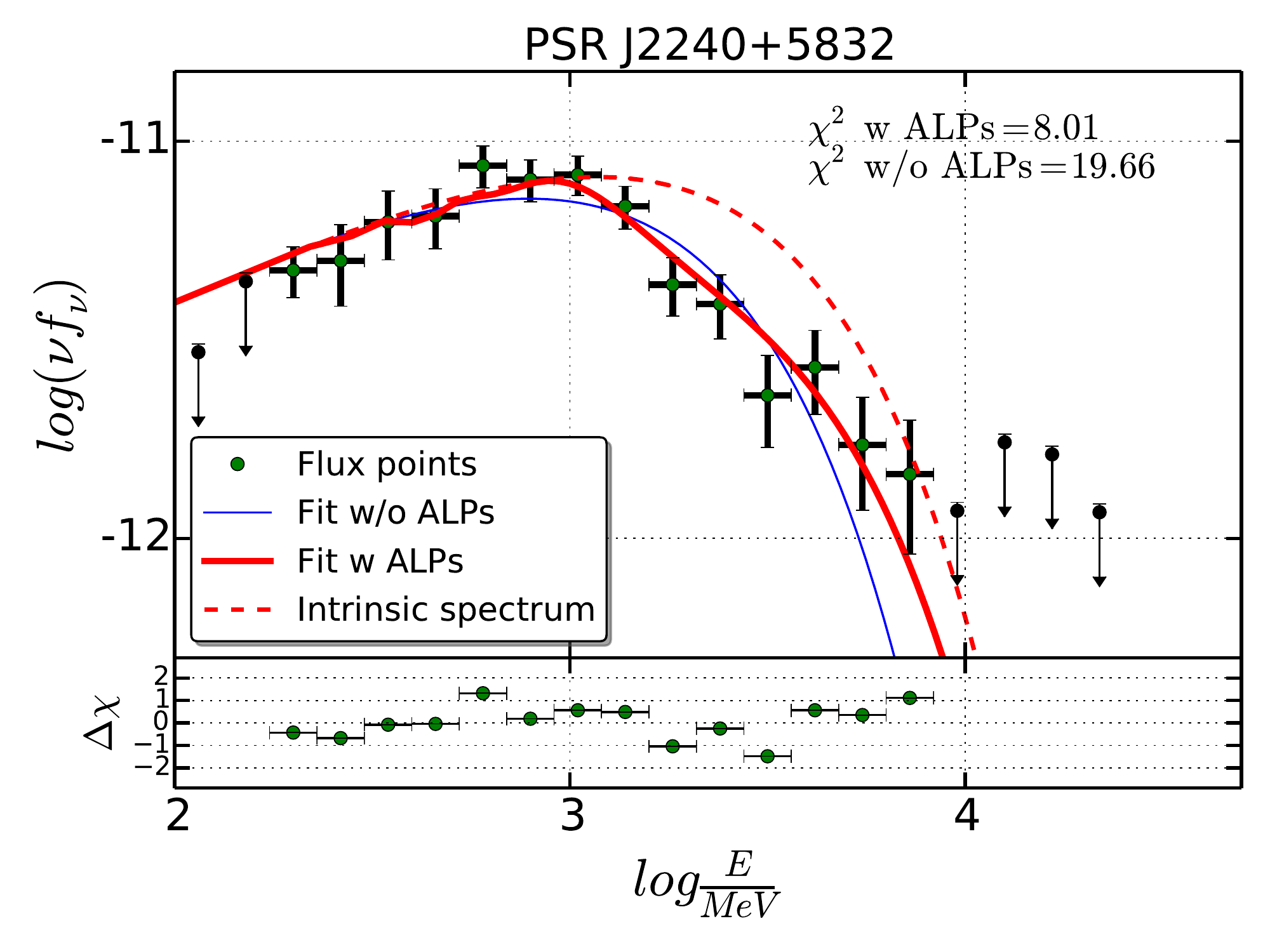}}
 \put(6.2,0){\includegraphics[width=8.6cm]{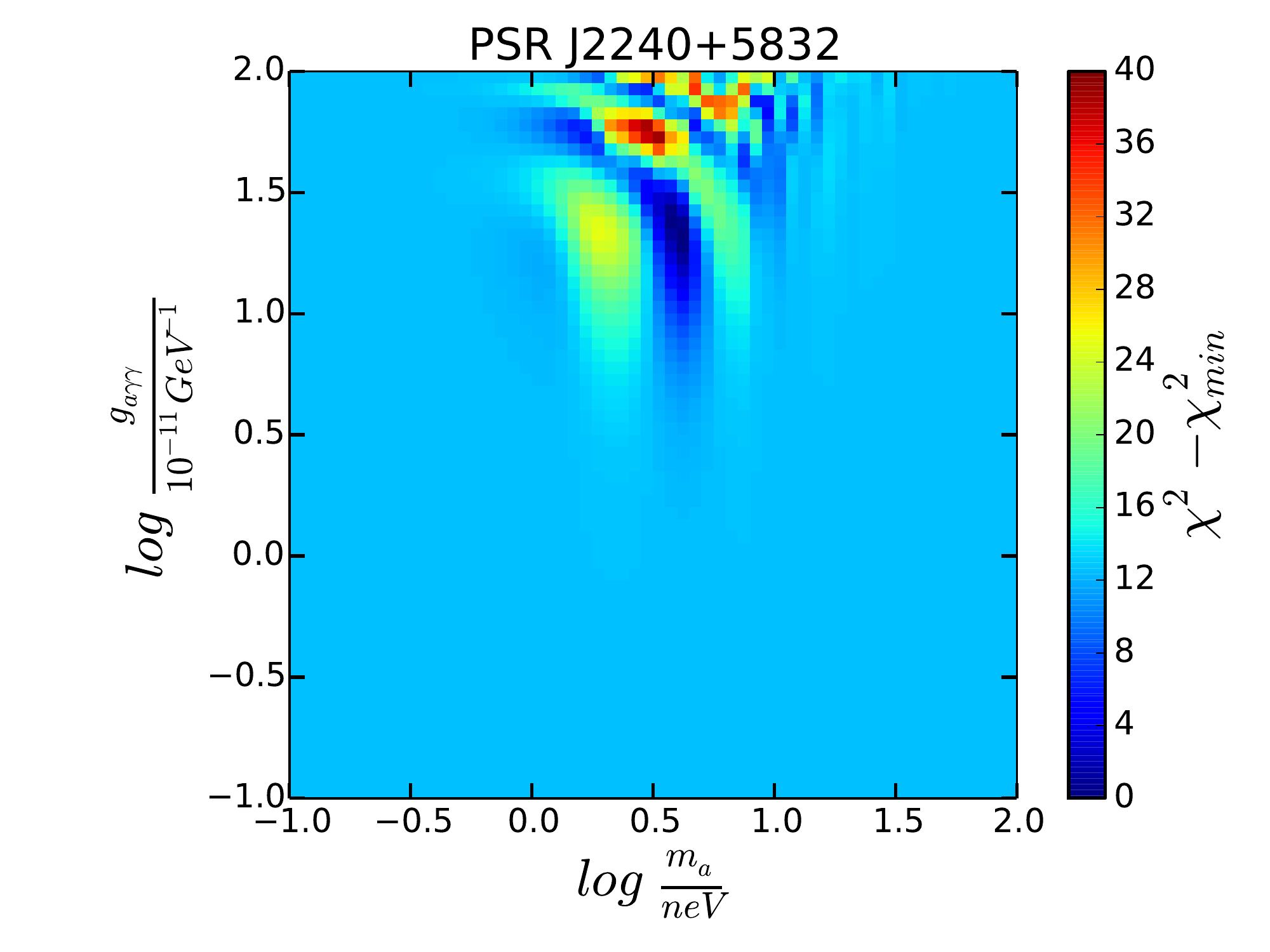}}
 \put(-3.2,7.5){\includegraphics[width=8.6cm]{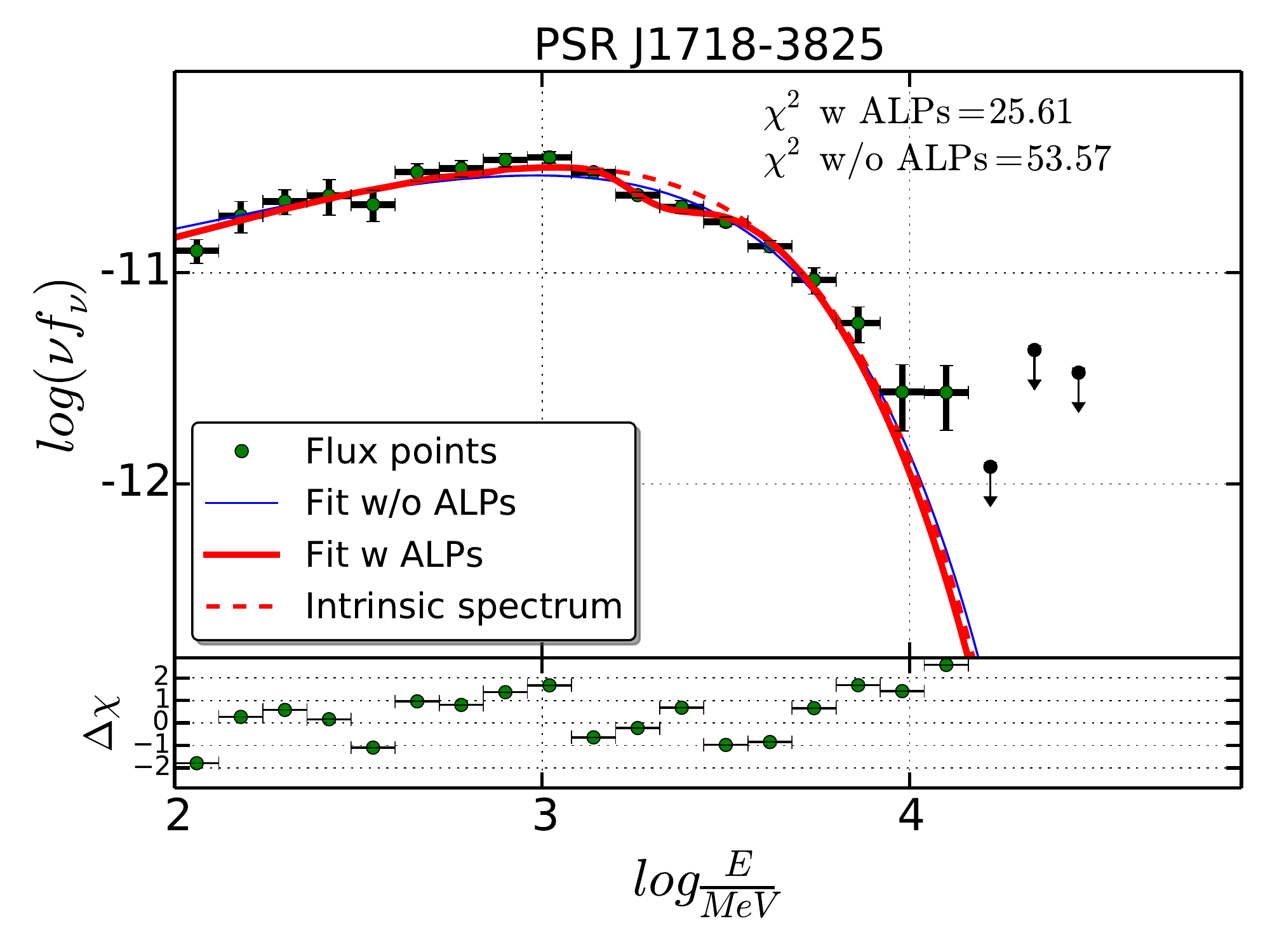}}
 \put(6.2,7.5){\includegraphics[width=8.6cm]{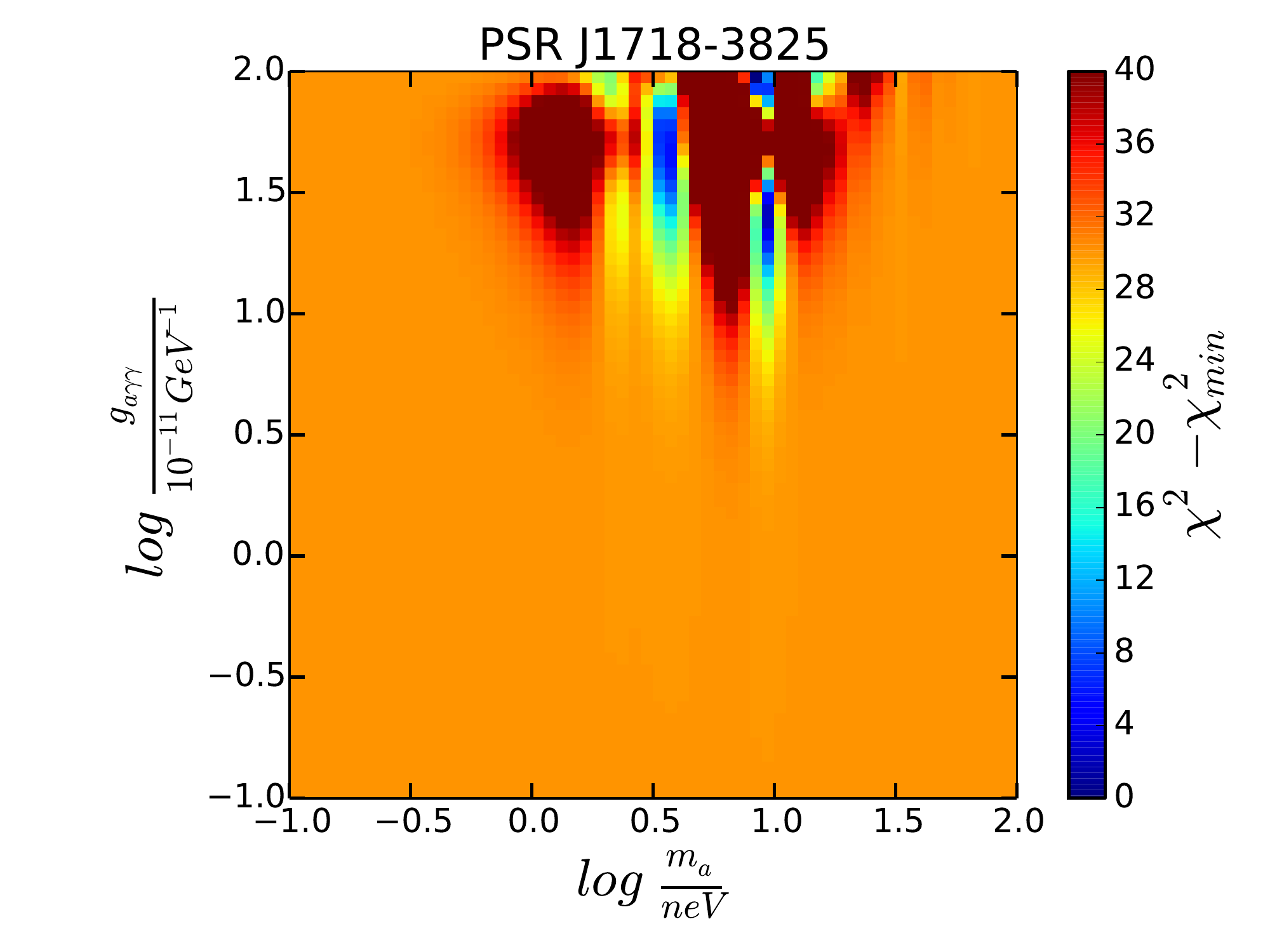}}
 
 \end{picture}
\end{center}
\caption{Same as Fig~\ref{fig:J1420_J1648} with different pulsar sources.
\label{fig:J1702_J2240}}
\end{figure}

The spectra of the analyzed objects are shown together with the best fitting models and the plane of parameters (figure~\ref{fig:J1420_J1648}: PSR J1420-6048  PSR~J1648-4611; figure~\ref{fig:J1702_J2240}: PSR~J1702-4128, PSR J1718-3825,  and PSR  J2240+5832). 

\newpage

\pagebreak

 \subsection{Photon-ALPs contour dependence on magnetic field parameters and the distance to the source}
\begin{figure}
\setlength{\unitlength}{.9cm}
\begin{center}
\begin{picture}(10,4)
 \put(-4.8,0){\includegraphics[width=8.5cm]{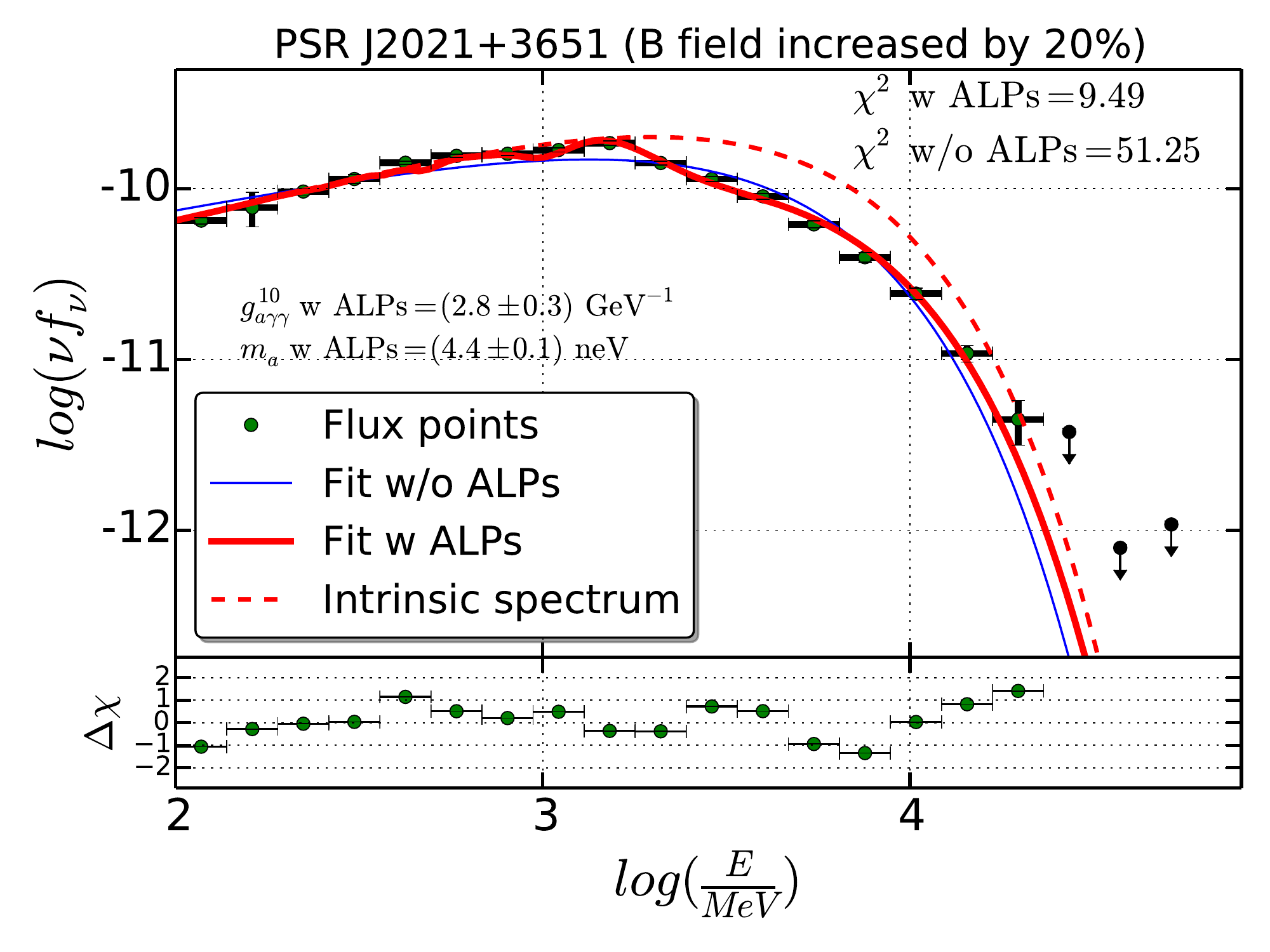}}
 \put(4.8,0){\includegraphics[width=8.5cm]{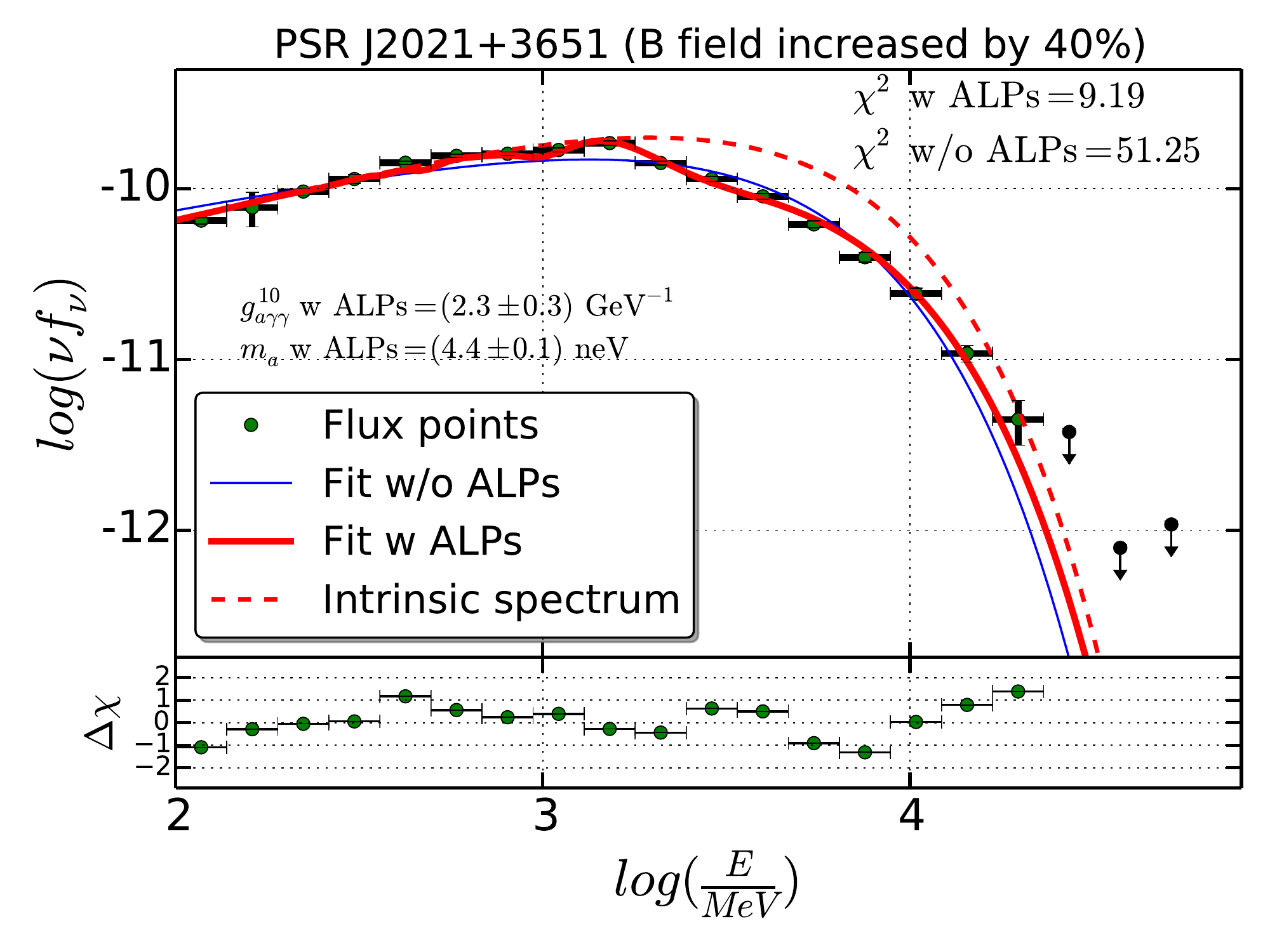}}
 \end{picture}
\end{center}
\caption{Variation of $g_{a\gamma\gamma}$ and $m_{a}$ with the change in Galactic magnetic field intensity. Left panel: The ALPs parameters are derived if we increase the magnetic field intensity by 20\%. Right Panel: the fitting corresponds to the magnetic field intensity increased by 40\% which reduces the   $g_{a\gamma\gamma}$ by 33.8\% whereas $m_{a}$ remains the same. 
	(note, $g_{a\gamma\gamma}^{10}$ is given in units of $10^{-10}$~GeV$^{-1}$)
\label{fig:Bfield_uncer_spectrum}}

\end{figure}

\begin{figure}
\setlength{\unitlength}{.9cm}
\begin{center}
\begin{picture}(10,10)
 \put(-4.8,0){\includegraphics[width=8.5cm]{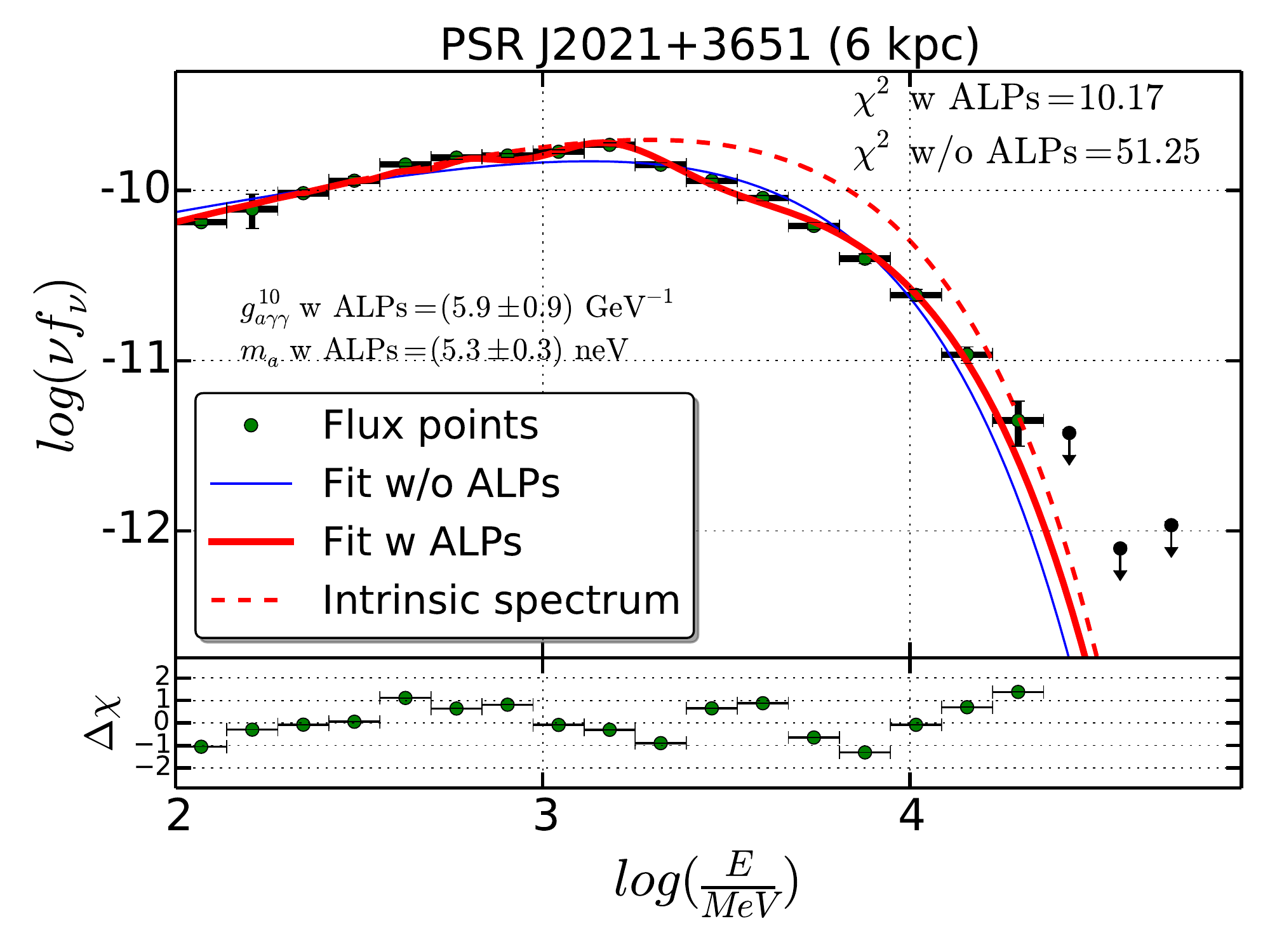}}
 \put(4.8,0){\includegraphics[width=8.5cm]{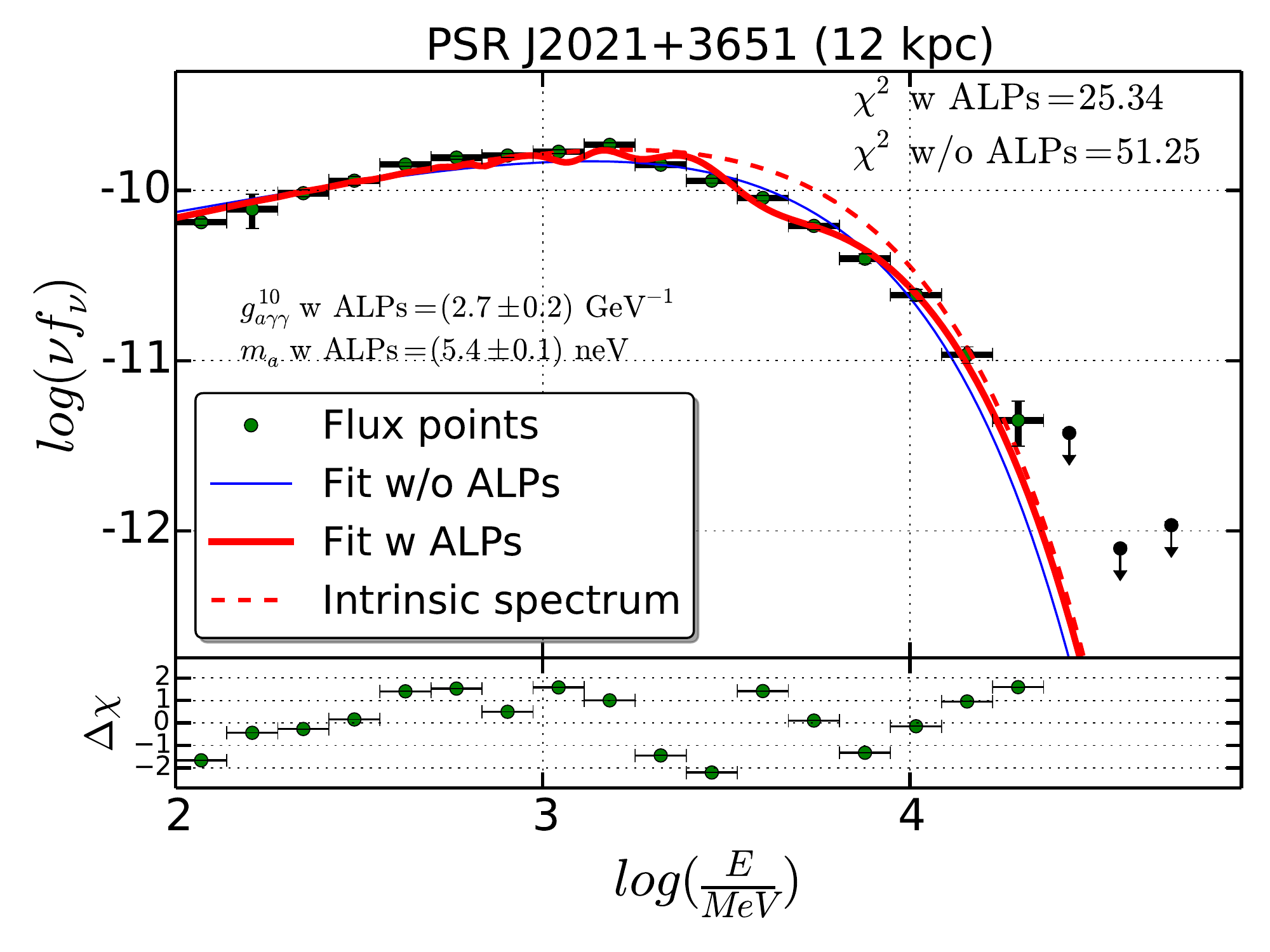}}
 \end{picture}
\end{center}
\caption{Pulsar spectra with the variation in the dintance to the source. In the left: pulsar spectrum are derived for a distance of 6 kpc while, in the right panel: the spectrum correspond to the distance of  12 kpc respectively.  
(note, $g_{a\gamma\gamma}^{10}$ is given in units of $10^{-10}$~GeV$^{-1}$)
\label{fig:distance_uncer_spectrum}   }

\end{figure}

 We already have discussed about the dependence of our best fit contour on Galactic magnetic field parameters and the source position in the section 4. The corresponding spectral fits are shown in  figures~\ref{fig:Bfield_uncer_spectrum} -- \ref{fig:distance_uncer_spectrum}.

\pagebreak

%
%
%

\end{document}